\documentclass[preprint,12pt]{elsarticle}

\usepackage{amssymb}
\journal{Physics Letter A}

\biboptions{sort&compress}

\usepackage{geometry}
\usepackage{fleqn}

\usepackage{graphicx} 
\usepackage{array}
\usepackage{bm}
\usepackage{color}
\usepackage{float}
\usepackage{subfigure}
\usepackage{multirow}
\usepackage{amsmath,amsfonts,amssymb}
\usepackage[colorlinks=true, allcolors=blue]{hyperref}

\begin{document}

\begin{frontmatter}

\title{Fast Evolution of Single Qubit Gate in Non-Adiabatic Geometric Quantum Computing} 

%

\author{Ge Tang, Xiao-Yong Yang}
\address{Department of Physics, Shanghai University, 200444 Shanghai, China}

\author{Ying Yan\corref{cor1}}
\ead{yingyan@suda.edu.cn}
\address{School of Optoelectronic Science and Engineering $\&$ Collaborative Innovation Center of Suzhou Nano Science and Technology, Soochow University, Suzhou 215006, China, Soochow University, 215006 Suzhou, China\\
Key Lab of Advanced Optical Manufacturing Technologies of Jiangsu Province $\&$  Key Lab of Modern Optical Technologies of Education Ministry of China\\
Engineering Research Center of Digital Imaging and Display, Ministry of Education, Soochow University}

\author{Jie Lu\corref{cor2}}
\ead{lujie@shu.edu.cn}
\address{Department of Physics, Shanghai University, 200444 Shanghai, China\\
 Shanghai Key Lab for Astrophysics, 100 Guilin Road, 200234 Shanghai, China}

\cortext[cor1]{Corresponding author}
\cortext[cor2]{Corresponding author}

\date{\today}

\begin{abstract}
  We implemented arbitrary single qubit gates of geometric quantum computing for a three-level system in a single-shot manner. The evolution time of the gate has been minimized by considering the shortest trajectory of the state on the Bloch sphere. The duration of gates grows from zero with the rotation angle  $\gamma$, and the tested T gate time can be reduced to $\sim$40\%   of those in the  traditional orange-sliced-shaped path non-adiabatic holonomic quantum computing (NHQC) scheme by the parametrization of Rabi frequency. We also demonstrated that certain pulses are robust against static detuning errors and Rabi errors.
The time-dependent detuning and Rabi frequency are found to be proportional to each other by a constant which is determined by the geometric phase. In this way, some previous NHQC schemes can be treated as special cases in our generalized model. 
\end{abstract}

%


\end{frontmatter}




\section{Introduction}
\label{sec:intro}
Quantum computing has become one of the most promising applications of quantum mechanics since 1990s, when novel algorithms has been proposed to tackle certain particular problems, such as integer factorization \cite{Shor1994} and unstructured database searching \cite{Grover1997}. A lot of efforts have been put in theoretical investigations  and experimental realizations during last decades. While the actual form of general quantum computing is yet to fully clear, a practical paradigm called noisy intermediate-scale quantum (NISQ) computers \cite{Preskill2018} is already available for quantum circuits with 50-100 qubits. A universal set of quantum circuits can be constructed using general single qubit gates and Controlled-NOT gates. Therefore, a key issue to achieve the NISQ computing is to lower down the infidelity of these two types of gates, which should be about from $10^{-6}$ to $10^{-2}$ depending on the error correction schemes \cite{Knill2005}, otherwise the stored information can't be transferred or retrieved effectively.

A single qubit gate can be realized by a phase shifting in an appropriate basis $|\phi_n\rangle\to e^{i\gamma_n}|\phi_n\rangle$, which turns into a general $SU(2)$ rotation $U(\bm{n},\gamma)$  in the two-dimensional Hilbert space in the computational space $\{|0\rangle, |1\rangle\}$. Usually, this transformation is implemented by a quantum evolution of laser(microwave)-matter interaction. The phase $\gamma_n$ can be geometric only if the dynamical phase is removed. So it is called geometric quantum computing (GQC) \cite{Johansson2012,Jones2000,Solinas2004,Solinas2012,Zanardi1999} or holonomic quantum computing (HQC)\cite{Xu2012,ZanardiRasetti2000, Zanardi2001,Erik2012,Duan2001}. Due to the global properties of the geometric phase, the GQC or HQC are inherently robust against small perturbations in the evolution trajectories of the qubit, such as decay, decoherence and systematic errors \cite{Johansson2012,Thomas2011, Zheng2016,Jing2017}. More details and references can be found in pedagogical papers \cite{Erik2015R,Erik2016R} and the recent thorough review on GQC and HQC \cite{ZhangJiang2021}.

Among many factors which can affect the fidelity, the duration of the operation is a crucial one.  The gate time needs to be decreased so that decay and decoherence effects which originated from the interaction between quantum system and environment would have less impact on the quantum state. That's the main reason why the adiabatic GQC scheme is not preferred since the evolution time must be long enough to satisfy the adiabatic condition. Therefore, the non-adiabatic geometric quantum computing (NGQC)\cite{Wang2001,Zhu2002} or the non-adiabatic Holonomic quantum computing (NHQC) \cite{Erik2012} has been dominated since they were proposed. 
 
The duration of single qubit gate is determined by the speed of quantum evolution and the length of the path on  the Bloch sphere. The general form of the time optimal solution has been investigated  \cite{Carlini2012,Carlini2013,Wang2015,Geng2016}, and it was applied in NHQC in superconducting qubit system \cite{BJLiu2020}. Another way to reduce the gate time is to shorten the length of the evolution path. Traditional NHQC schemes usually require exactly the same length of the evolution path for any angles of rotation, sometimes it is called orange-slice-shaped NHQC (OSS-NHQC) \cite{Cen2006,Feng2007,Wu2007,Kim2008,Feng2009,Chen2012,Zhao2016}. A systematic approach to minimize the length has been proposed for two-level \cite{Li2020,Ding2021}  and three-level systems \cite{Zhao2020,Liang2021}. In this work, we minimized the path of a single qubit gate in a three-level system with a $\Lambda$ configuration in the framework of NHQC. Our scheme naturally reduces to the original NHQC \cite{Erik2012} with $\gamma=\pi$, and to the single-shot NHQC with constant detuning \cite{Erik2016}.
The gate duration can be further shortened by  increasing the maximal value of Rabi frequency, which is determined by a parameter $k$ available in the pulses.
 
This paper is organized as follows. We first introduce the concept of the geometric phase and NHQC in section \ref{sec:NHQC}; then we describe the detailed modelling for NHQC based on a thre-level system with the $\Lambda$ configuration in section \ref{sec:three_level}. In section \ref{sec:path_mini}, we propose a new method for   minimizing the length of the path which corresponds to the circular passage on the Bloch sphere, and our method can be reduced to some traditional NHQC schemes  under certain conditions. In section \ref{sec:simulation},  we numerically analyse the relation of the gate duration $\tau$ and rotation angle $\gamma$, then simulate the gate robustness against the detuning error and the Rabi error with the Lindblad master equation.  At last, we provide the discussion and conclusion in section \ref{sec:conclusion}.

\section{Geometric phase}
\label{sec:NHQC}
In order to perform geometric quantum computing, we first need to construct  a time-dependent orthonormal basis $\{|\phi_n(t)\rangle\}$. Each of them satisfies the Schr$\rm{\ddot{o}}$dinger equation\footnote{We take the convention of $\hbar=1$ in this work.}
\begin{equation}
i\frac{d}{dt}|\phi_n(t)\rangle=H(t)|\phi_n(t)\rangle
\label{sch_eq}
\end{equation} 
in the Hilbert space $\mathcal{H}$. In this way the transitions between different  $|\phi_n(t)\rangle$  have been forbidden by eq.(\ref{sch_eq}) without restriction of  the adiabatic condition. As a special case, this evolution reduces to  an adiabatic process if $|\phi_n(t)\rangle$ happens to be the instantaneous eigenstates of $H(t)$. Sometimes $\{|\phi_n(t)\rangle\}$ is called the dynamical basis in literatures. One considers the cyclic evolution, where the initial and final state of $|\phi_n(t)\rangle$ only differ by a phase factor $\gamma_n(t)$ as 
\begin{equation}
|\phi_n(\tau)\rangle=e^{i\gamma_n(t)} |\phi_n(0)\rangle ,
\label{cyclic_phi}
\end{equation}
where $\tau$ is the duration of  the evolution. In this case the physical state evolves back to  the original one $|\phi_n(\tau)\rangle\langle\phi_n(\tau)|=|\phi_n(0)\rangle\langle\phi_n(0)|$. Now we define a new basis $\{|\chi_n(t)\rangle\}$ in  the projective Hilbert space $\mathcal{P(H)}$
\begin{equation}
|\chi_n(t)\rangle=e^{-i\gamma_n} |\phi_n(t)\rangle .
\label{chi_phi}
\end{equation}
The cyclic evolution condition is then 
\begin{equation}
|\chi_n(\tau)\rangle = |\chi_n(0)\rangle = |\phi_n(0)\rangle .
\label{cyclic_chi}
\end{equation}
The total  phase factor $\gamma_n$ includes both the dynamical part $\gamma^d_n$ and  the geometric part $\gamma^g_n$, which  are defined as 
\begin{align}
\gamma^d_n &= -\int^\tau_0 dt\langle \phi_n(t)|H(t)| \phi_n(t)\rangle, \label{gamma_d}\\ 
\gamma^g_n &= i\int^\tau_0 dt\left\langle \chi_n(t)\left|\frac{d}{dt}\right| \chi_n(t)\right\rangle,
\label{gamma}  
\end{align}
where the non-adiabatic Abelian geometric phase $\gamma^g_n$ is also called the Aharonov-Anandan (A-A) phase, and it becomes the Berry phase in the limit of adiabatic evolution. 

Generally, there are two options to remove the dynamical phase. One is imposing the parallel transport condition
\begin{equation}
\langle \phi_n(t)|H(t)| \phi(t)_n\rangle = 0,
\label{parallel_condition}
\end{equation}
therefore the dynamical phase is zero at any time. The other is that the dynamical phase only vanishes at the end of evolution,
\begin{equation}
\int^\tau_0\langle \phi_n(t)|H(t)| \phi_n(t)\rangle dt= 0,
\end{equation}
 e.g., utilizing methods like the spin-echo technique. In any cases, the general time-evolution operator in the N-dimensional Hilbert space is 
\begin{equation}
U(\tau)=\sum^N_{n,m} \mathcal{T}\exp\left\{i\int^\tau_0\left[A(t)-H(t)\right]\right\}_{mn} |\phi_n(\tau)\rangle\langle \phi_m(0)|,
\label{Utau_nonabelian}
\end{equation}
where $\mathcal{T}$ is the time-ordering operator, $A_{nm}(t)=i\langle\chi_n(t)|\dot{\chi}_m(t)\rangle$, and $H_{mn}(t)=\langle\phi_m(t)|H(t)|\phi_n(t)\rangle$.  Utilizing the conditions (\ref{cyclic_phi})-(\ref{cyclic_chi}) and (\ref{parallel_condition}), the above formula can be written as
\begin{equation}
U(\tau)=\sum_n e^{i\gamma_n} |\phi_n(0)\rangle\langle \phi_n(0)|,
\label{Utau_phi}
\end{equation}
where $\gamma_n=\gamma^g_n$ now is the geometric phase shift on $|\phi_n(0)\rangle$.

\section{NHQC on three-level system}
\label{sec:three_level}
\begin{figure}[H]
\begin{minipage}{8cm}
\centering
	\includegraphics[width=0.5\textwidth]{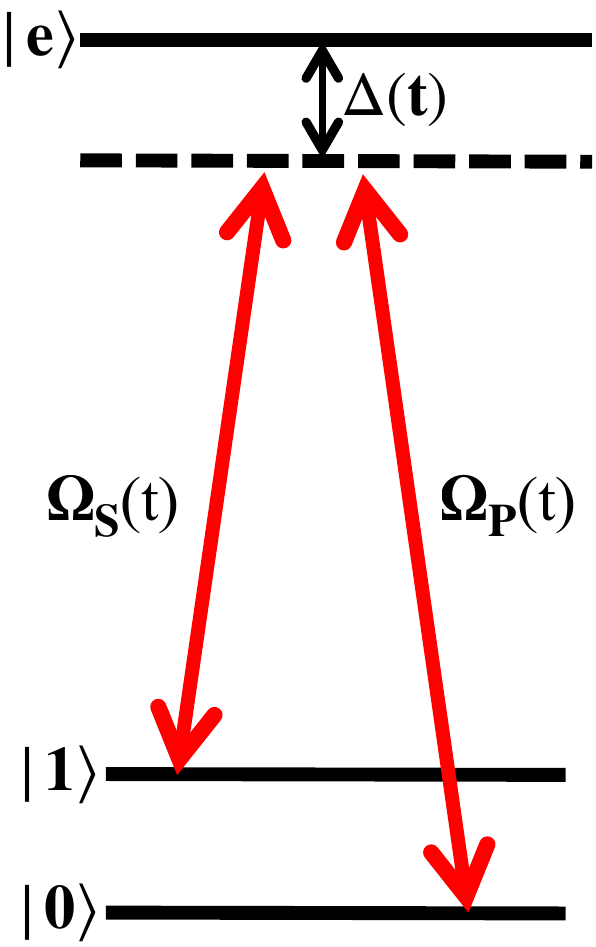} %
\centerline{(a)}
\end{minipage}
\hfill  
\begin{minipage}{8cm}
\centering
	\includegraphics[width=0.5\textwidth]{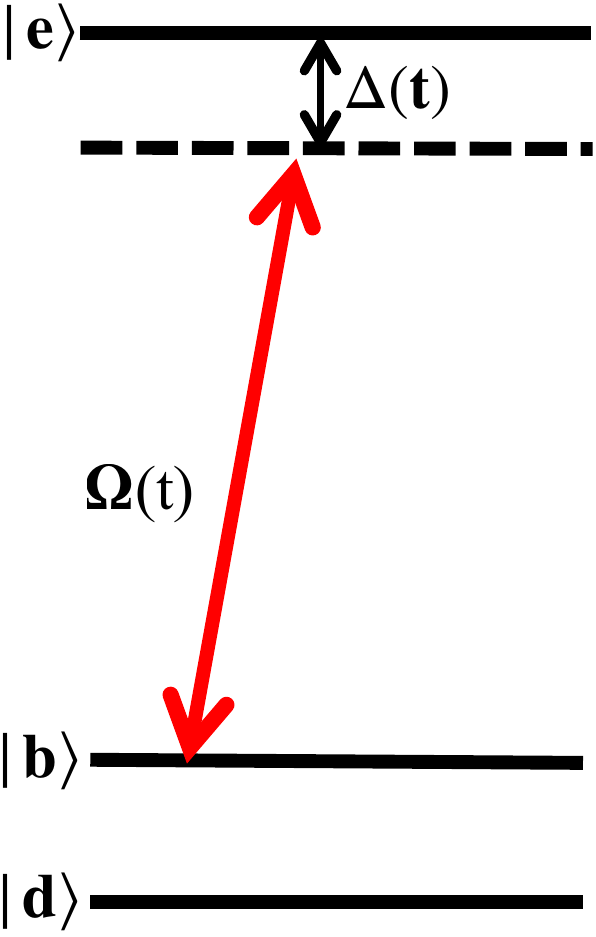}
\centerline{(b)}
\end{minipage}
	\caption{The schematic figures of a three-level system. (a) A  three-level system with $\Lambda$ configuration in the basis of $\{|0\rangle,|1\rangle,|e\rangle\}$, the transitions $|0\rangle-|e\rangle$ and $|1\rangle-|e\rangle$ are driven by the pump and Stokes pulses. (b) The effective two-level system in the basis of $\{|b\rangle,|e\rangle\}$ where the dark state $|d\rangle$ is decoupled.}
\label{fig:3level}
\end{figure}

We consider a three-level system with the $\Lambda$ configuration, as depicted in Figure \ref{fig:3level}, a pair of laser(microwave) pulses with central frequency $\nu_{S,P}(t)$ interacting with the system.  Under the rotating wave approximation in the interaction picture, and transforming to the rotating frame, the Hamiltonian of this system can be written as 
\begin{equation}
H(t)=\Delta(t)|e\rangle\langle e|+[\Omega_{P}(t)|0\rangle \langle e|+\Omega_{S}(t)|1\rangle\langle e|+ \rm{h.c.}]\ ,\label{3level_Hamiltonian}
\end{equation}
where the time-dependent detuning is defined as $\Delta(t)=2\pi\nu_{S,P}(t)-\omega_{ei}$, and $\omega_{ei}=E_e-E_i$ ($i=0,1$). The pump and Stokes pulse are respectively parametrized as
\begin{equation}
\Omega_{P}(t)=\Omega(t)\sin\frac{\theta}{2}e^{-i\xi(t)},\\
\end{equation} 
and 
\begin{equation}
\Omega_{S}(t)=-\Omega(t)\cos\frac{\theta}{2}e^{i[\phi-\xi(t)]},
\end{equation} 
where $\Omega(t)$ is the Rabi frequency, and $\phi$ is the relative phase  between the two pulses.

The time-independent bare basis $\{|0\rangle,|1\rangle, |e\rangle\}$ forms a three-dimensional Hilbert space $\mathcal{H}$.  The initial state vector is parametrized in the computational subspace of  $\{|0\rangle,|1\rangle\}$ as
\begin{equation}
|\psi(0)\rangle=\cos\theta_0|0\rangle+\sin\theta_0 e^{i\phi_0}|1\rangle.
\label{psi0}
\end{equation} 
where the $\theta_0$ and $\phi_0$ are  angles in the range of $[0,\pi]$ and $[0,2\pi]$, respectively. Therefore the qubit is encoded in the computational subspace of  $\{|0\rangle,|1\rangle\}$. 
At the end of evolution, the state vector $|\psi(\tau)\rangle$ evolves back to the same subspace with desired populations and phases in $|0\rangle$ and $|1\rangle$, equivalently rotated by an $SU(2)$ operator, i.e. 
\begin{equation}
|\psi(0)\rangle \to |\psi(\tau)\rangle =  U(\bm{n},\gamma) |\psi(0)\rangle.
\end{equation}
$|e\rangle$ is the auxiliary state with zero population at $t=0$ and $\tau$.

It is well noted that there is a "dark state" $|d\rangle$ for Hamiltonian (\ref{3level_Hamiltonian}) with eigenvalue of zero, i.e. $H(t)|d\rangle=0$, and an orthogonal "bright state" $|b\rangle$ as follows
\begin{align}
|d\rangle &=\cos\frac{\theta}{2}\left|0\right\rangle +\sin\frac{\theta}{2}e^{i\phi}\left|1\right\rangle \label{eq:dark},\\
|b\rangle &=\sin\frac{\theta}{2}\left|0\right\rangle -\cos\frac{\theta}{2}e^{i\phi}\left|1\right\rangle .
\label{eq:bright}
\end{align}
Thus the  subspace $\{|d\rangle,|b\rangle\}$ is equivalent to  the computational subspace $\{|0\rangle,|1\rangle\}$.
With the definition (\ref{eq:dark}) and (\ref{eq:bright}), the Hamiltonian (\ref{3level_Hamiltonian}) can be  rewritten as an effective two-level system which oscillates between  the bright state $|b\rangle$ and  the auxiliary state $|e\rangle$
\begin{align}
\tilde{H}(t)=\Delta(t)|e\rangle\langle e|+\left[e^{-i\xi(t)}\Omega(t)|b\rangle\langle e|+\text{h.c}\right],
\label{Hamiltonian_dbe}
\end{align}
where the dark state has been decoupled, as depicted in Figure \ref{fig:3level}(b). Similar to the normal two-level system, we can construct an orthonormal basis $\{|\chi_{+}(t)\rangle,|\chi_{-}(t)\rangle\}$ as a linear combination of $|b\rangle$ and $|e\rangle$
\begin{align}
|\chi_{+}(t)\rangle &=\cos\frac{\alpha(t)}{2}|b\rangle +\sin\frac{\alpha(t)}{2}e^{i\beta(t)}\left|e\right\rangle, \\
|\chi_{-}(t)\rangle &=\sin\frac{\alpha(t)}{2}|b\rangle -\cos\frac{\alpha(t)}{2}e^{i\beta(t)}\left|e\right\rangle.
\end{align}
These are the basis of the 2-dimensional projective Hilbert space, just as eq.(\ref{chi_phi}).

\begin{figure}[H]
\centering
\includegraphics[width=1.0\textwidth]{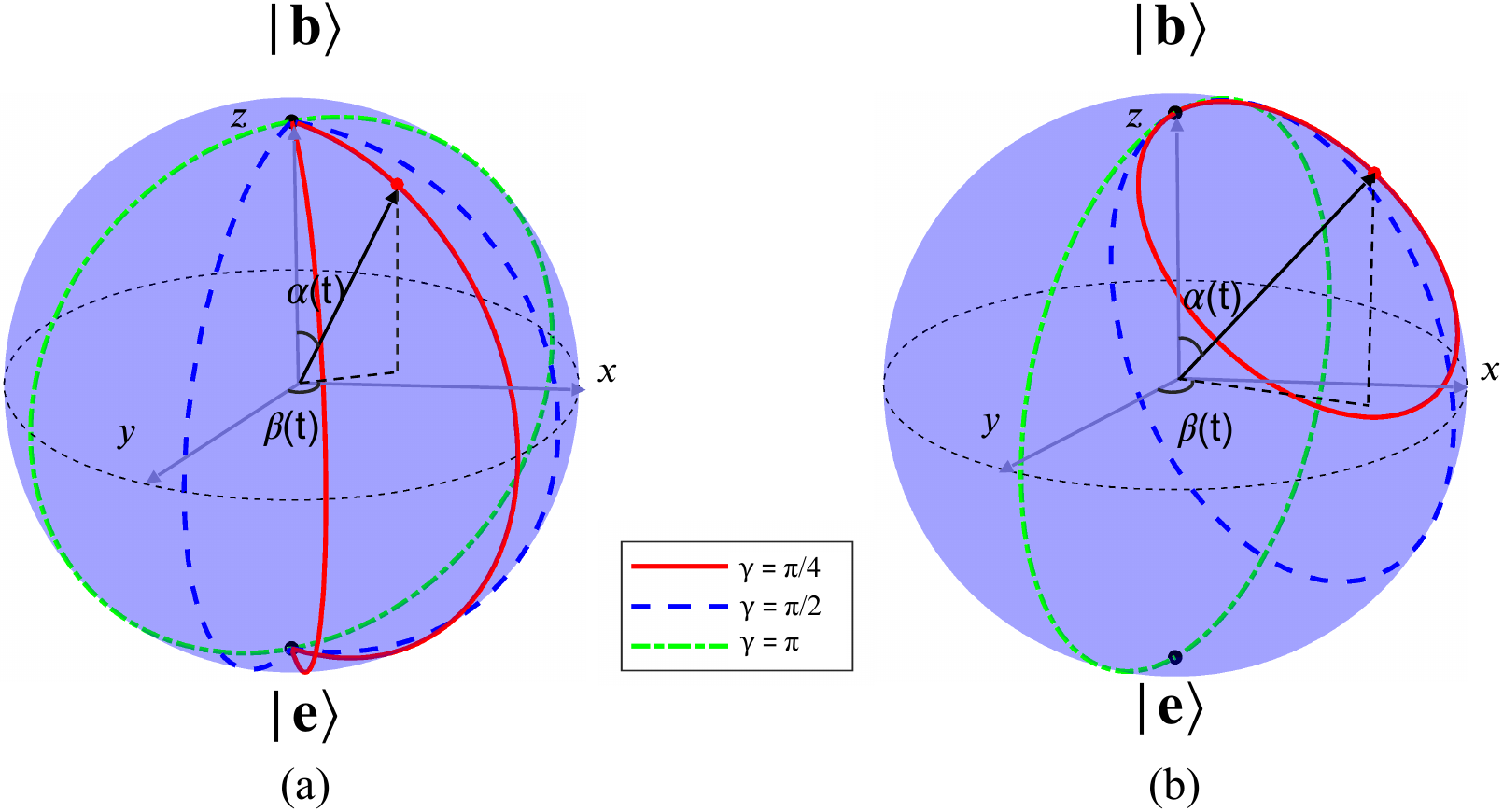} %
\caption{The illustrative paths on  {\color{blue}the} Bloch sphere: (a) NHQC with Orange sliced shaped paths; (b) NHQC with circular paths. {\color{blue}The two black dots represent the state $|b\rangle$ and  $|e\rangle$ respectively, and the red dot represent the state $|\chi_{+}(t)\rangle$.} }
\label{fig:Bloch_sphere}
\end{figure} 

A traditional recipe to implement geometric gate is OSS-NHQC \cite{Cen2006,Feng2007,Wu2007,Kim2008,Feng2009,Chen2012,Zhao2016}. The $\alpha(t)$ and  $\beta(t)$ can be set as 
\begin{align}
\alpha(t)&=\pi\sin^2\left(\frac{\pi t}{\tau}\right), \label{alpha_NHQC}\\
\beta(t) &= \begin{cases}
\beta_1,\quad 0\leq t <\tau/2\\
\beta_2,\quad \tau/2 \leq t \leq \tau \label{beta_NHQC}\\
\end{cases}.
\end{align} 
The cyclic evolution condition (\ref{cyclic_chi}) for $|\chi_+(t)\rangle$ becomes
\begin{equation}
\alpha(0)=\alpha(\tau)=0 .
\label{cyclic_alpha}
\end{equation}  
Therefore the state  $|\chi_+(t)\rangle$ evolves from the northern pole $|b\rangle$ to the southern pole $|e\rangle$ with a constant  phase $\beta_1$, and evolves back to the northern pole $|b\rangle$ with another constant  phase $\beta_2$ in Figure \ref{fig:Bloch_sphere}(a), so the geometric phase $\gamma=\beta_1-\beta_2$. 

The same as eq.(\ref{sch_eq}), the physical pure state $\rho_\pm(t) =|\phi_\pm(t)\rangle\langle \phi_\pm(t)| = |\chi_\pm(t)\rangle\langle \chi_\pm(t)|$ should satisfy the Von Neumann equation
\begin{align}
i\frac{d\rho_\pm(t)}{dt}&=\left[\tilde{H}(t),\rho_\pm(t)\right],
\end{align}
which generates the following constraints
\begin{align}
\Omega(t) &= -\frac{\dot{\alpha}(t)}{2\sin[\beta(t)-\xi(t)]},
\label{Omegat} \\
\xi(t) &= \beta(t)-\arctan\left[\frac{\dot{\alpha}(t)}{\Delta(t)+\dot{\beta}(t)}\cot\alpha(t)\right]. 
\label{xi}
\end{align}
The dynamical phase can be eliminated by  the parallel transport condition (\ref{parallel_condition}), which results in 
\begin{equation}
\Delta(t)=-\dot{\beta}(t)[1+\cos\alpha(t)].
\label{parallel_delta}
\end{equation}
Therefore the  time-evolution operator (\ref{Utau_phi}) is
\begin{equation}
U(\tau)=|d\rangle\langle d|+e^{i\gamma}|b\rangle\langle b|,
\label{Utau_db_gamma}
\end{equation}
where the accumulated geometric phase
\begin{align}
\gamma &= i\int^\tau_0 \langle\chi_+(t)|\dot{\chi}_+(t)\rangle  dt\nonumber\\
       &= -\int^\tau_0 \dot{\beta}(t)\sin^2\frac{\alpha(t)}{2} dt.
\label{gamma}
\end{align}
In the computational subspace $\{|0\rangle,|1\rangle\}$, $U(\tau)$ becomes a general $SU(2)$ rotation
\begin{equation}
U(\tau)=e^{i\frac{\gamma}{2}}e^{-i\frac{\gamma}{2}(\bm{n}\cdot \bm{\sigma})},
\label{Utau_01_gamma}
\end{equation}
with  an extra global phase $e^{i\frac{\gamma}{2}}$ which can be ignored,  and $\bm{n}=(\sin\theta\cos\phi, \sin\theta\sin\phi, \cos\theta)$ is the rotation axis.

\section{Minimizing the path}
\label{sec:path_mini}

An apparent drawback of OSS-NHQC recipe is that all gates have exactly the same travelling length, regardless of their rotation angles. Therefore, various proposals have been proposed to shorten the path on the Bloch sphere in recent years \cite{GFXu2018,Li2020,Zhao2020,Ding2021,Liang2021, Wood2020},  where a simple geometric solution is to let the state trajectory be a circle, which is the minimal path for a given area on the Bloch sphere. This idea can be traced back to the work in 1988 \cite{Suter1988}.
 A circle passing through {\color{blue}the} northern pole can be parametrized as follows \cite{Zhao2020}
\begin{equation}
(\pi-\gamma)[1-\cos\alpha\left(t\right)]=\frac{\ell_{\rm{c}}}{2}\sin\alpha(t)\cos\beta(t)
\label{circle}
\end{equation}
where $\ell_{\rm{c}}=2\sqrt{2\pi\gamma-\gamma^{2}}$ is the circumference of the circle. The area enclosed by this trajectory is $2\gamma$. The magnitude of $\alpha(t)$ reaches the maximal value 
\begin{equation}
\alpha_{\rm m}(\gamma)=\arccos[(\pi^2-4\pi\gamma+2\gamma^2)/\pi^2]
\end{equation}
at $t=\tau/2$. Note $\beta(t)$ can be shifted by an arbitrary constant value, but the shape of circle keeps the same.
Combining the constraints in eq.(\ref{Omegat}, \ref{xi}, \ref{parallel_delta}, \ref{circle}), we find that $\xi(t)=0$ and equations (\ref{Omegat}) and (\ref{xi}) change to
\begin{align}
\Omega(t)&=
\frac{1}{2}\sqrt{\dot{\beta}^{2}(t)\sin^{2}\alpha(t)+\dot{\alpha}^{2}(t)},\label{Omega_parallel}\\
\Delta(t) &= -2\Omega(t)\frac{\pi-\gamma}{\sqrt{2\pi\gamma-\gamma^2}}. \label{Delta_parallel}
\end{align}
The Rabi frequency $\Omega(t)$ and detuning $\Delta(t)$  are proportional to each other with a constant, which means that the laser frequency has to be time-dependent for $\gamma\neq\pi$.

Now the Hamiltonian (\ref{Hamiltonian_dbe}) can be expressed as 
\begin{equation}
\tilde H(t)= \frac{1}{2}\Delta(t)\tilde{I}+\frac{\pi}{\sqrt{2\pi\gamma-\gamma^2}}\Omega(t) \tilde{\bm{n}}(\gamma)\cdot\tilde{\bm\sigma},
\end{equation}
where $\tilde{I}$ and $\tilde{\bm\sigma}$ are the identity and Pauli operators in the $\{|e\rangle,|b\rangle\}$ basis, and $\tilde{\bm{n}}(\gamma)=(\sqrt{2\pi\gamma-\gamma^2}/\pi,0,(\pi-\gamma)/\pi)$ is the time-independent roational axis.  Obviously, $\tilde H(t)$ commutes with itself at any two instant times, i.e., $[\tilde H(t_1),\tilde H(t_2)]=0$,  so the time-ordering operator $\mathcal{T}$ can be removed from the time-evolution operator. As a result, $\tilde{U}(\tau)$ in the $\{|e\rangle,|b\rangle\}$ subspace can be calculated as follows
\begin{align}
\tilde{U}(\tau)&=\exp\left[-i\int^\tau_0\tilde{H}(t)dt\right]\nonumber \\
&=\exp(-i\Phi)\left[I\cos\left(\Theta\over2\right)-\tilde{\bm{n}}(\gamma)\cdot \tilde{\bm{\sigma}}\sin\left(\Theta\over2\right)\right],
\label{Utau_be}
\end{align}
where 
\begin{align}
\Phi&=\int^\tau_0\frac{\Delta(t)}{2}dt,\label{eq:Phi}\\
\Theta&=2\int^\tau_0\frac{\pi\Omega(t)}{\sqrt{2\pi\gamma-\gamma^2}}dt.\label{eq:Theta}
\end{align}
The cyclic evolution condition (\ref{cyclic_chi}) requires $\Theta=2\pi$, which can be written as 
\begin{equation}
\int^\tau_0\Omega(t)dt=\frac{\ell_c}{2}.
\label{cyclic_omega}
\end{equation}
It shows a rotation of $2\pi$ around the time-independent axis $\tilde{n}(\gamma)$ in the $\{|e\rangle,|b\rangle\}$ basis, and the travelling length of state $|\chi_+(t)\rangle$ is $\ell_c$. 

In the limit of $\gamma=\pi$, detuning $\Delta=0$ according to eq.(\ref{Delta_parallel}), and $\tilde{\bm{n}}(\gamma)$ becomes the $x$ axis.  This is exactly the case of the original NHQC in Ref.\cite{Erik2012} and the cyclic evolution condition (\ref{cyclic_omega}) becomes
$\int^\tau_0 \Omega(t)dt=\pi$. The same condition also holds for the single shot OSS-NHQC, except that the time-independent axis $\tilde{\bm{n}}(\gamma)$ has to be different in the  two intervals, as we can see from Figure \ref{fig:Bloch_sphere}(a) and eq.(\ref{beta_NHQC}).

If the dark state $|d\rangle$ is included, the time-evolution operator in the subspace $\{|d\rangle, |b\rangle\}$ under the cyclic evolution condition is 
\begin{equation}
U(\tau)=|d\rangle\langle d|-e^{-i\Phi}|b\rangle\langle b|.
\label{Utau_Phi}
\end{equation}
Comparing to eq.(\ref{Utau_db_gamma}), the geometric phase $\gamma$ can be expressed as 
\begin{align}
\gamma &=\pi-\Phi \nonumber\\
&=\frac{1}{2}\int_{0}^{\tau}\left[1-\cos\alpha(t)\right]\dot{\beta}(t)dt
\label{gamma_Phi}
\end{align}
As a special case, if the detuning $\Delta$ is a constant, the Rabi frequency must be a square pulse $\Omega(t)=\Omega_0$ according to eq.(\ref{Delta_parallel}). Thus the time-evolution operator reduces to the case in Ref.\cite{Erik2016,Liu2021}, and the gate duration is
\begin{align}
\tau &= \frac{2\pi}{\sqrt{\Delta^2+4\Omega_0^2}} = \frac{\sqrt{2\pi\gamma-\gamma^2}}{\Omega_0}.
\label{eq:tau_gamma_constant}
\end{align}
If $\gamma=\pi$, this duration further reduces to $\tau=\pi/\Omega_0$, which is the case in Ref.\cite{Erik2012} with square pulses. We can clearly see that $\tau$ only depends on the geometric phase $\gamma$ when $\Omega_0$ is fixed.

\section{Numerical simulation}
\label{sec:simulation}
Besides the constraints shown in eq.(\ref{cyclic_alpha}) and (\ref{circle}), one more is needed to finalize the Rabi frequency $\Omega(t)$. 
We propose the azimuth angle $\alpha(t)$ as follows
\begin{align}
\alpha(t) &= \alpha_m(\gamma)\bar{\alpha}(t)^{k+1},
\label{alpha_1}
\end{align}
where $k$ is a positive real number and $\bar{\alpha}(t)$ is a  parabolic function as
\begin{align}
\bar{\alpha}(t)= 4\left(\frac{t}{\tau}-\frac{t^2}{\tau^2}\right).
\label{alpha_bar}
\end{align}
One usually prefers the Rabi frequency $\Omega(0)=\Omega(\tau)=0$ in experiments, that is guaranteed by $k>0$ according to eq.(\ref{Omegat}). The polar angle $\beta(t)$ is thus determined by eq.(\ref{circle}), and $\beta(\tau)-\beta(0)=\pi$.

\begin{figure}[H]
\centering
\includegraphics[width=0.8\textwidth]{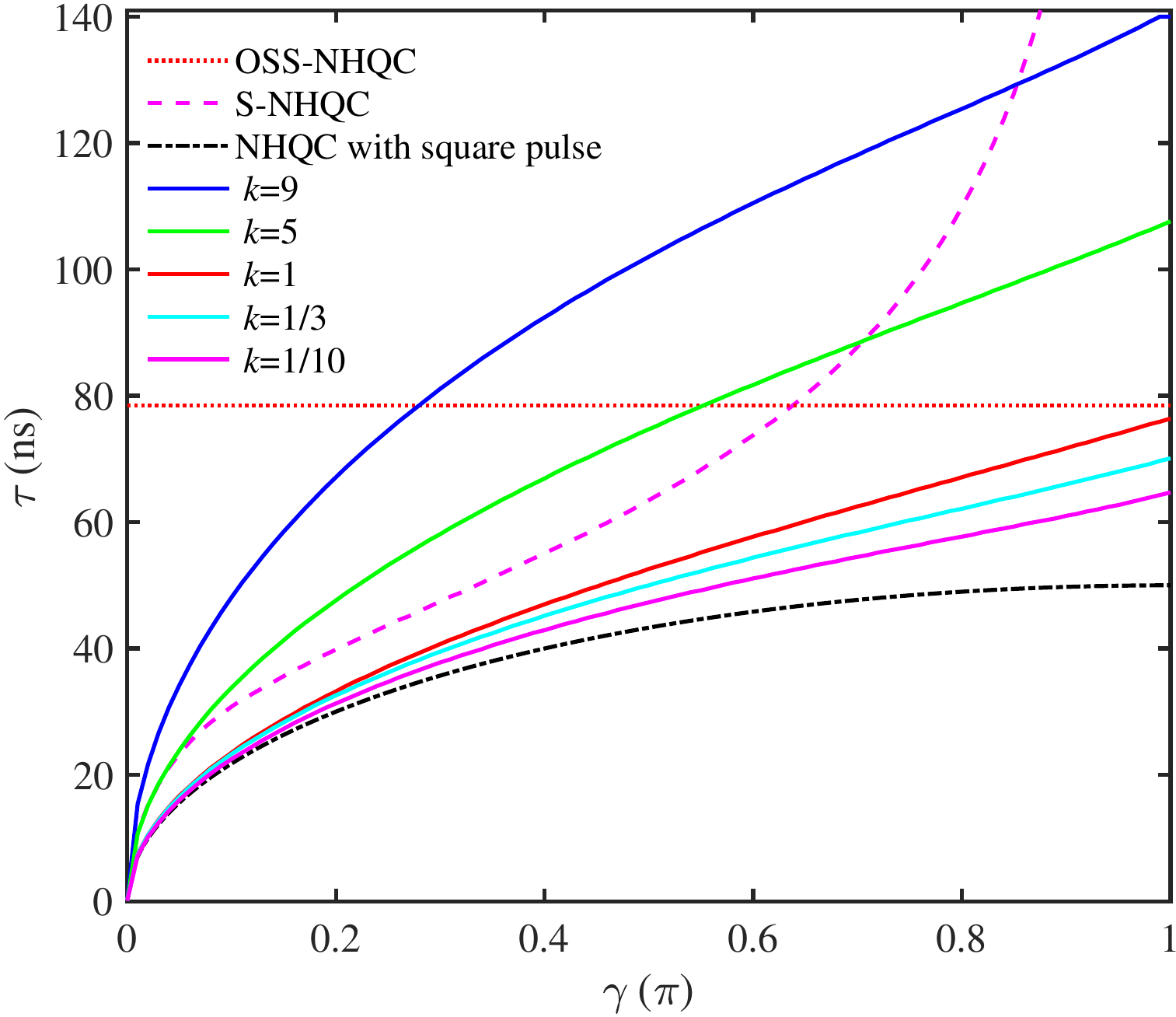}
	\caption{Dependence of  the gate duration $\tau$ on the geometric phase $\gamma$. The solid-red line indicates the OSS-NHQC recipe, the solid-blue line the S-NHQC scheme,  and the solid-green line the NHQC with squared pulse. Our scheme is shown with different $k$ values.}
\label{fig:gamma_tau}
\end{figure}

With the constraints in eq.(\ref{circle}-\ref{Omega_parallel}) and (\ref{alpha_1}), the relation between $\tau$ and $\gamma$ of (\ref{eq:tau_gamma_constant}) can be generalized to  an arbitrary time-dependent Rabi frequency $\Omega(t)$. We plotted the dependence of  the  gate operation time $\tau$ on the geometric phase $\gamma$ with different values of $k=1/10,\,1/3,\,1,\,5,\,$ and $9$.  Results are shown in Figure \ref{fig:gamma_tau}, where all maximal value of $\Omega(t)$ have been fixed to $\Omega_{\rm m}=2\pi\times10$ MHz. It is shown that the gates with larger $k$ value have longer duration for the same $\gamma$. The reason is that $\alpha(t)$ with larger $k$ has  a steeper slop, which leads to a shaper shape of $\Omega(t)$. In order to maintain the cyclic condition (\ref{cyclic_omega}) while keeping the same $\Omega_{\rm m}$, the gate duration $\tau$ has to be longer. 

The Rabi frequency of  the OSS-NHQC scheme is given by eq.(\ref{alpha_NHQC}-\ref{beta_NHQC}) and (\ref{Omegat}), and the maximum  $\Omega(t)$ is set to the same $\Omega_{\rm m}$ as mentioned above. In this case, the gate duration $\tau$ always keeps 78.5 ns regardless  of the value of $\gamma$, as shown by dotted-red line. A similar method to ours is the S-NHQC scheme \cite{Liang2021}, which also use the method of the circular path, whose parametrization is as follows
\begin{align}
\beta(t) &= \pi\sin^2\left(\frac{\pi t}{2\tau}\right),\\
\alpha(t)&= 2\arctan\left[\frac{\sqrt{2\pi\gamma-\gamma^2}}{\pi-\gamma}\sin\beta(t)\right].
\end{align}
This scheme cannot reach the limit $\gamma=\pi$, and it is shown by the dashed-magenta line in Figure \ref{fig:gamma_tau}.
It's clear that the gate duration $\tau$ of our scheme is shorter than those in the other two schemes for the same $\gamma$, provided that the $k$ is small. The square pulse scheme in dash-dotted-black has even shorter $\tau$, which is given by eq.(\ref{eq:tau_gamma_constant}). Its evolution time reaches $\tau=50$ ns at $\gamma=\pi$ if $\Omega_0=\Omega_{\rm m}$. This can also be understood by the cyclic evolution condition eq.(\ref{cyclic_omega}), where the integration of $\Omega(t)$ is a fixed value for a specific $\gamma$. Therefore, the gate duration of square pulse scheme can be regarded as the lower bound of this method. However, the square pulse is not preferred in many experiments due to multiple reasons \cite{Yan2019}. 

In many quantum computing system, there are two major errors arising from experimental operations, which are the detuning error $\Delta(t)\to \Delta(t)+\delta$ and the Rabi error $\Omega(t)\to (1+\epsilon)\Omega(t)$.  The evolution passage will be shifted away from the exact circular trajectory eq.(\ref{circle}) with errors, and eventually affect the fidelity of gates. The quantum evolution of system is evaluated by the Lindblad master equation
\begin{equation}
\frac{d}{dt}\rho(t)=i[\rho(t),H(t)]+\frac{1}{2}[\Gamma_{1}\mathcal{L}\left(\sigma_{1}\right)+\Gamma_{2}\mathcal{L}\left(\sigma_{2}\right)],
\label{eq:Lindblad}
\end{equation}
where $\mathcal{L(A)}=2\mathcal{A}\rho\mathcal{A}-\mathcal{A}^{\dagger}\mathcal{A}\rho-\rho\mathcal{A}^{\dagger}\mathcal{A}$ is the Lindbladian operator, $\sigma_{1}=|0\rangle \langle e|+\sqrt{2}|e\rangle \langle 1|+\sqrt{3}|1\rangle\langle h|$, $\sigma_{2}=|e\rangle \langle e|+2|1\rangle \left\langle 1\right|+3|h\rangle \langle h|$ with $|h\rangle $ indicating the third excited state, $\Gamma_{1}$ and $\Gamma_{2}$ are the corresponding decay and dephasing rates. Here, we consider the case $\Gamma_{1}=\Gamma_{2}=3\times2\pi$ kHz in superconducting circuits. The fidelity is defined as
\begin{equation}
F=\langle \psi_{\rm tg}|\rho(\tau)|\psi_{\rm tg}\rangle ,
\label{eq:fidelity}
\end{equation}
where $|\psi_{\rm tg}\rangle$ denotes the target state.

\begin{figure}[H]
\begin{minipage}{8cm}
\centering
\includegraphics[width=7.5cm,height=5cm]{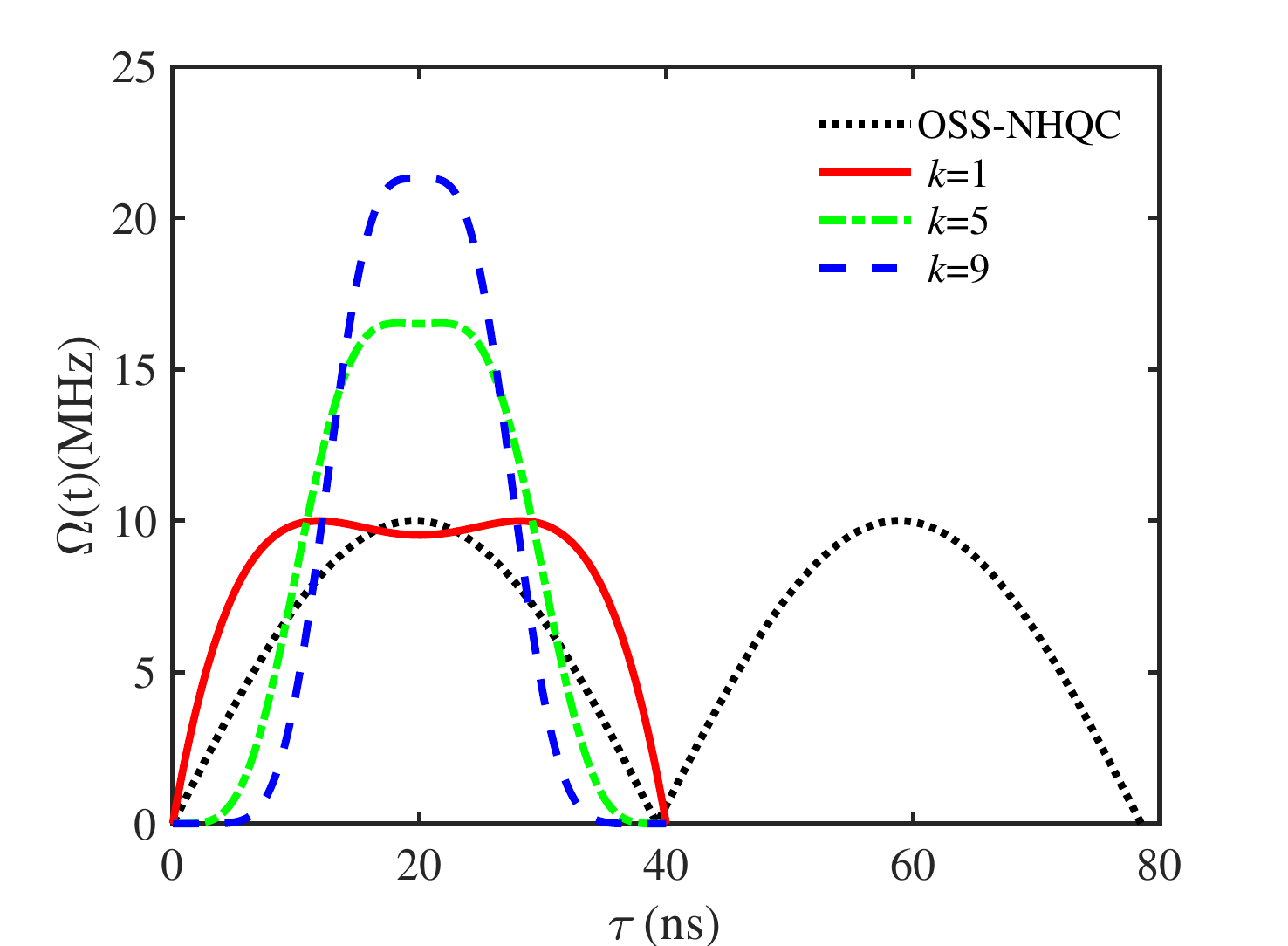}
\centerline{(a) T gate}
\end{minipage}
\hfill  
\begin{minipage}{8cm}
\centering
\includegraphics[width=7.5cm,height=5cm]{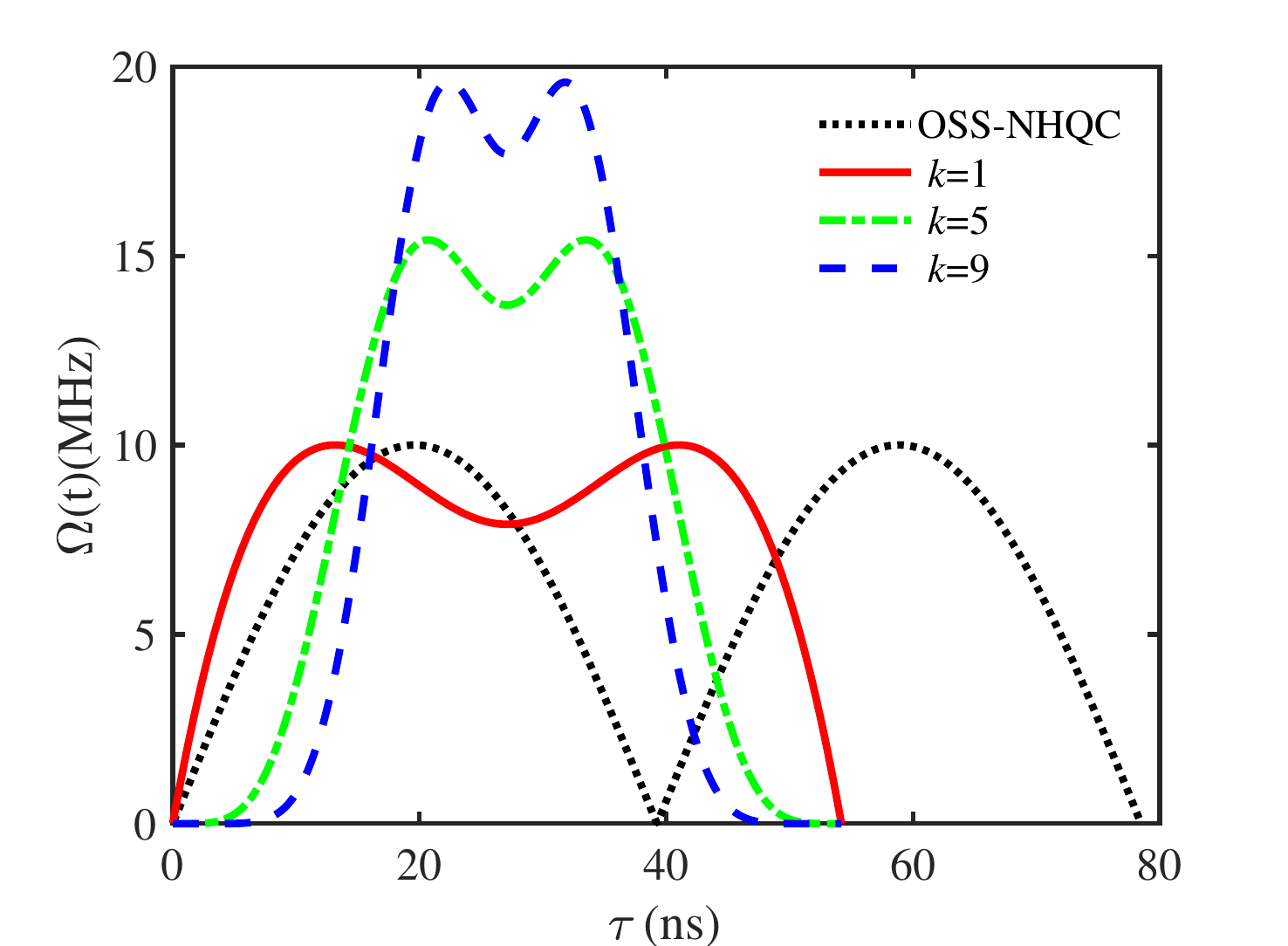} %
\centerline{(b) S gate}	
\end{minipage}
\vfill
\begin{minipage}{8cm}
\centering
	\includegraphics[width=7.5cm,height=5cm]{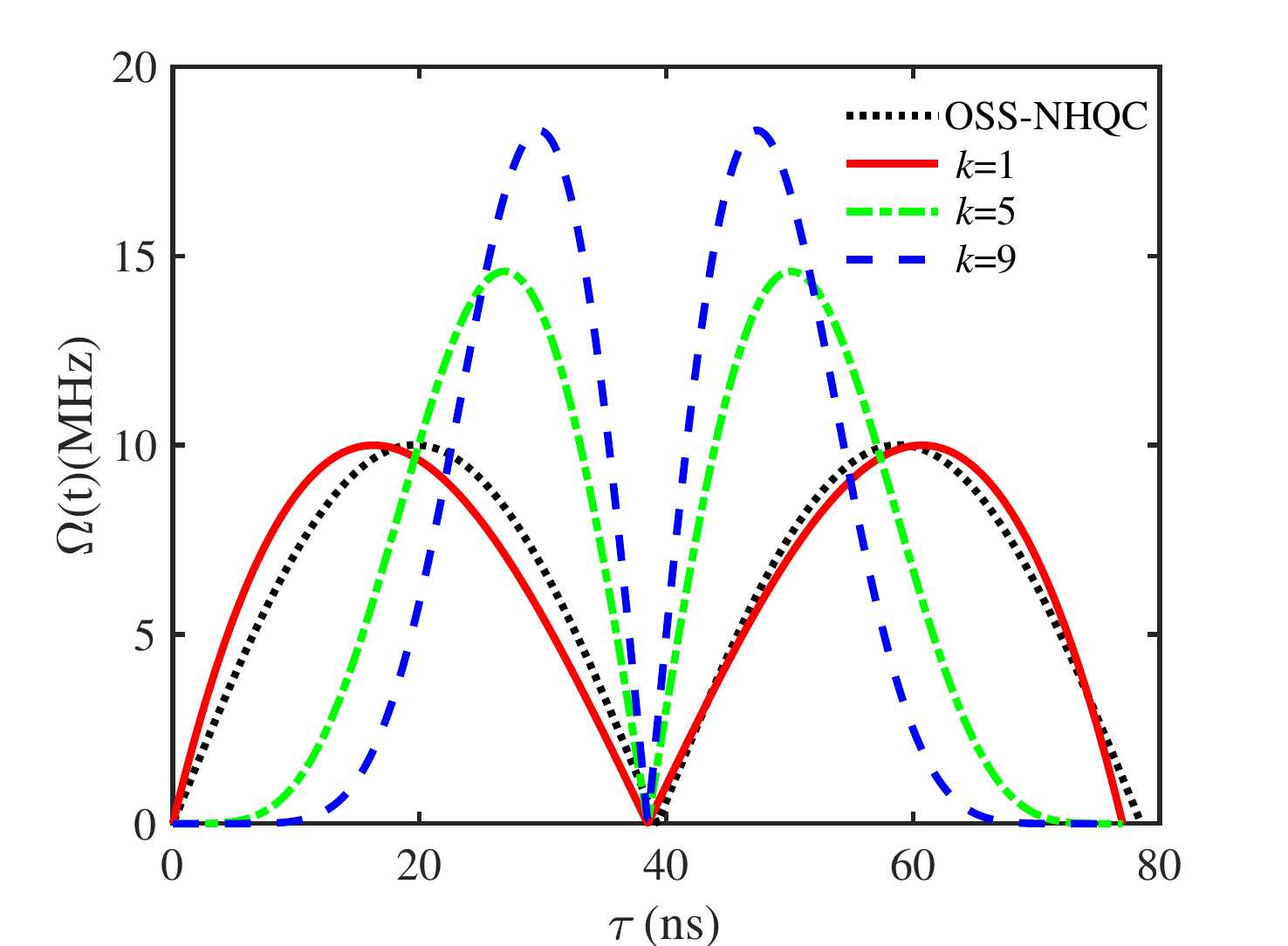}
\centerline{(c) Not gate}
\end{minipage}
\hfill
\begin{minipage}{8cm}
\centering
	\includegraphics[width=7.5cm,height=5cm]{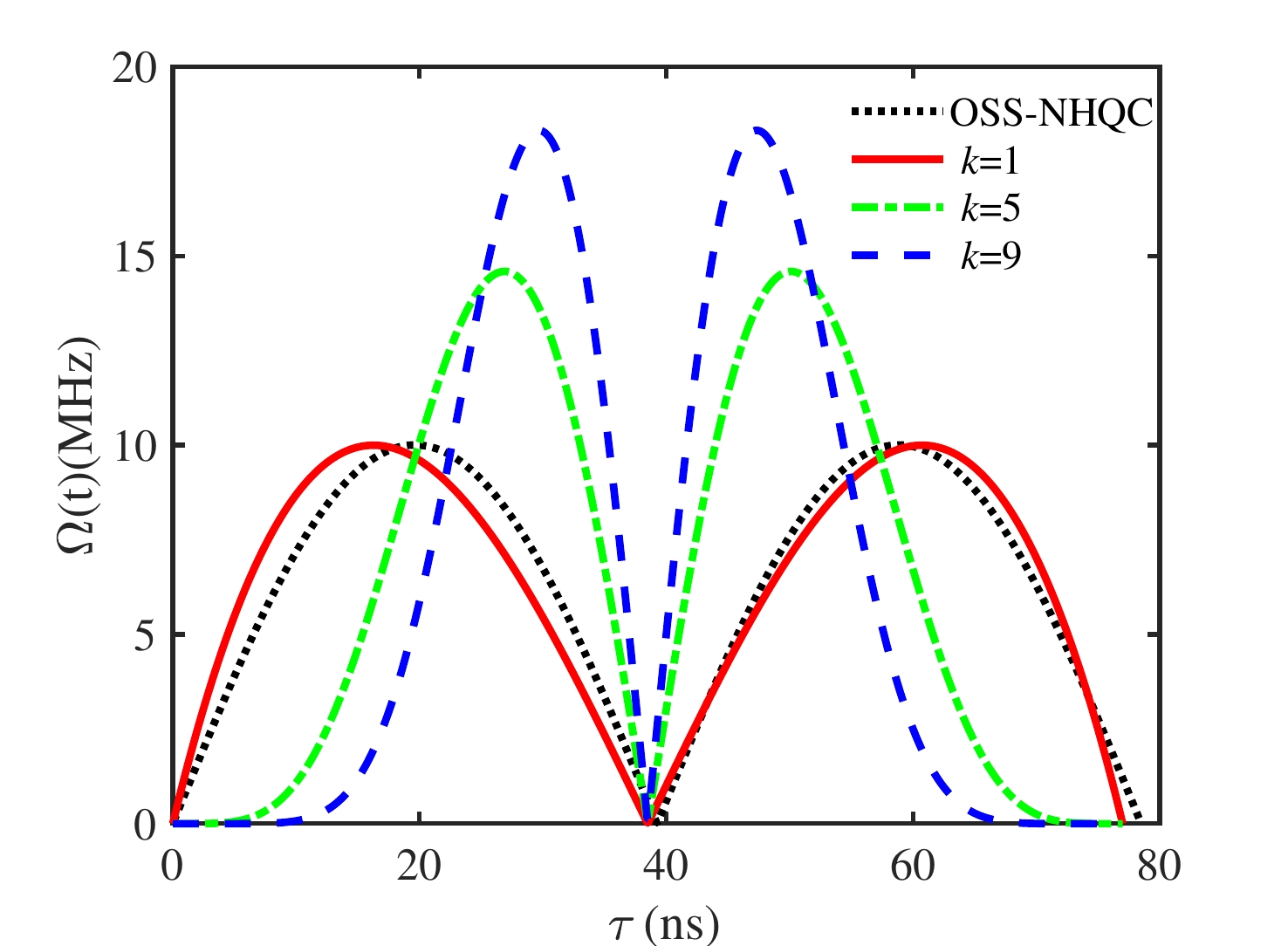} %
\centerline{(d) Hadamard gate}
\end{minipage}
\caption{(Color online) Envelopes of the Rabi frequency $\Omega(t)$ of pulses with different $k$ in the OSS-NHQC and our scheme for (a) T gate, (b) S gate, (c) Not gate, and (d) Hadamard gate. }
	\label{fig:gate_pulse}
\end{figure}

We investigated four different gates in the numerical simulation, T gate $(\gamma=\pi/4,\theta=\phi=0)$, S gate $(\gamma=\pi/2,\theta=\phi=0)$, NOT gate $(\gamma=\pi,\theta=\pi/2, \phi=0)$ and Hadamard gate $(\gamma=\pi,\theta=\pi/4,\phi=0)$. From previous discussion, we know that either the $\Omega_{\rm m}$ or the $\tau$ shall be fixed for a particular pulse. In order to take the same measure  of the de-coherence effects in the Lindblad equation, we fixed the gate duration for all the $k$ values for a specific gate. Therefore, the gate duration are determined by the case $k=1$, which are 39.8 ns for T gate, 54.2 ns for S gate and 77.0 ns for Not gate and Hadamard gate. The height of Rabi frequency $\Omega(t)$ is thus determined by the value of $k$, as we can see from Figure \ref{fig:gate_pulse}. The pulse envelops for $k=1,5,9$ and  the OSS-NHQC are plotted for four gates. 

\begin{figure}[H]
\begin{minipage}{8cm}
\centering
	\includegraphics[width=7.5cm,height=5cm]{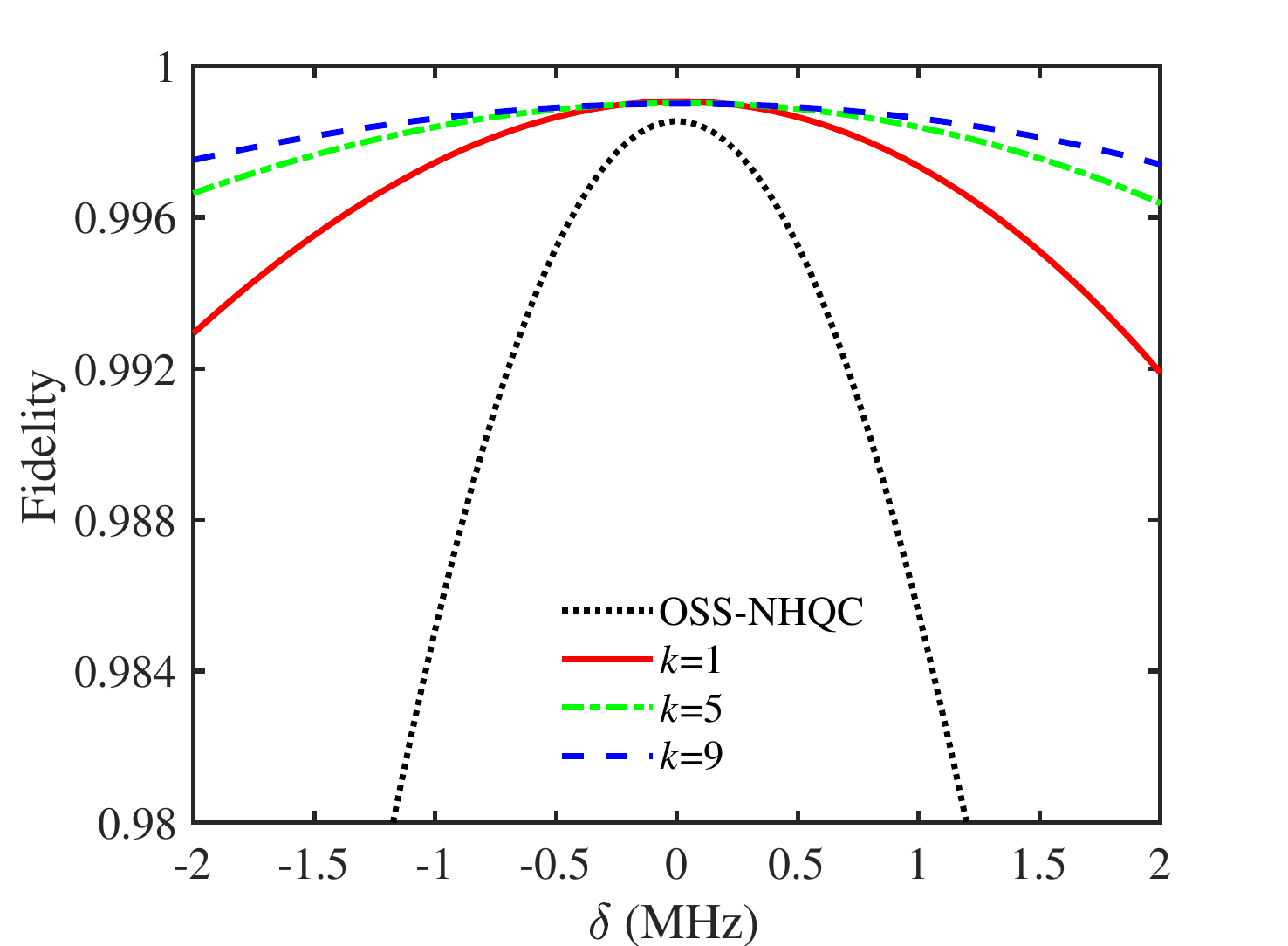} %
\centerline{(a) T gate }
\end{minipage}
\hfill  
\begin{minipage}{8cm}
\centering
	\includegraphics[width=7.5cm,height=5cm]{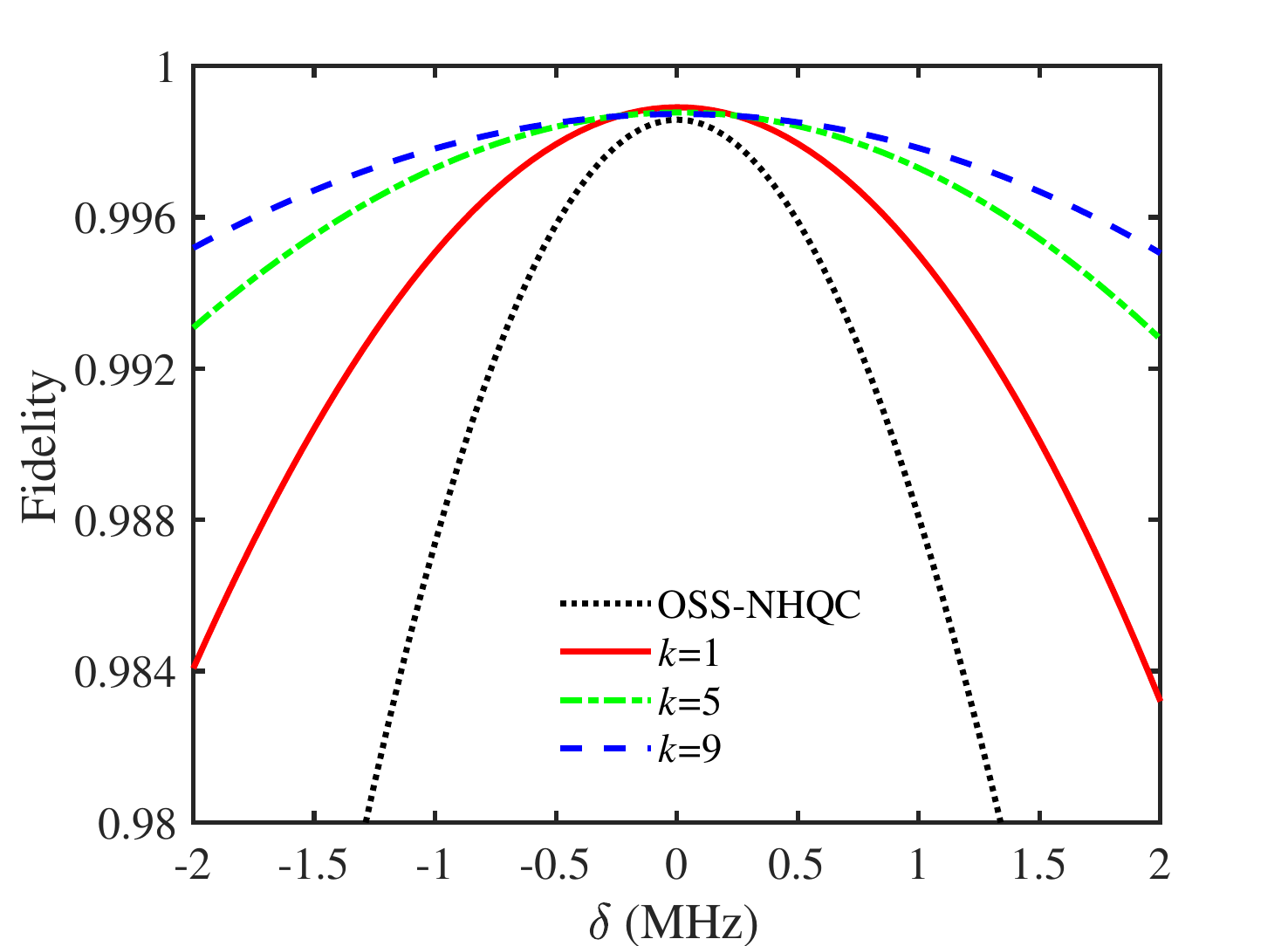}
\centerline{(b) S gate}
\end{minipage}
\vfill
\begin{minipage}{8cm}
\centering
	\includegraphics[width=7.5cm,height=5cm]{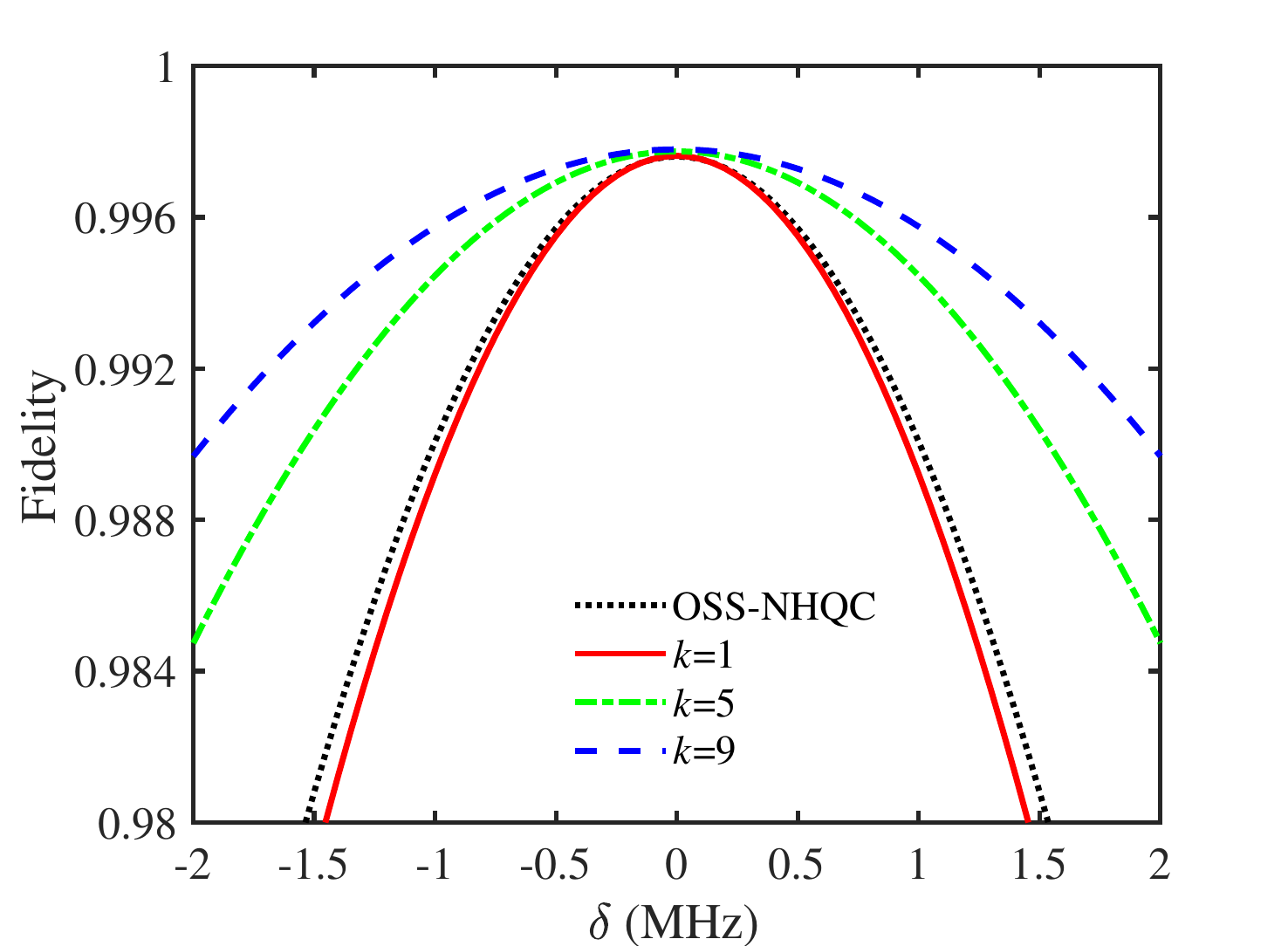}
\centerline{(c) Not gate}
\end{minipage}
\hfill
\begin{minipage}{8cm}
\centering
	\includegraphics[width=7.5cm,height=5cm]{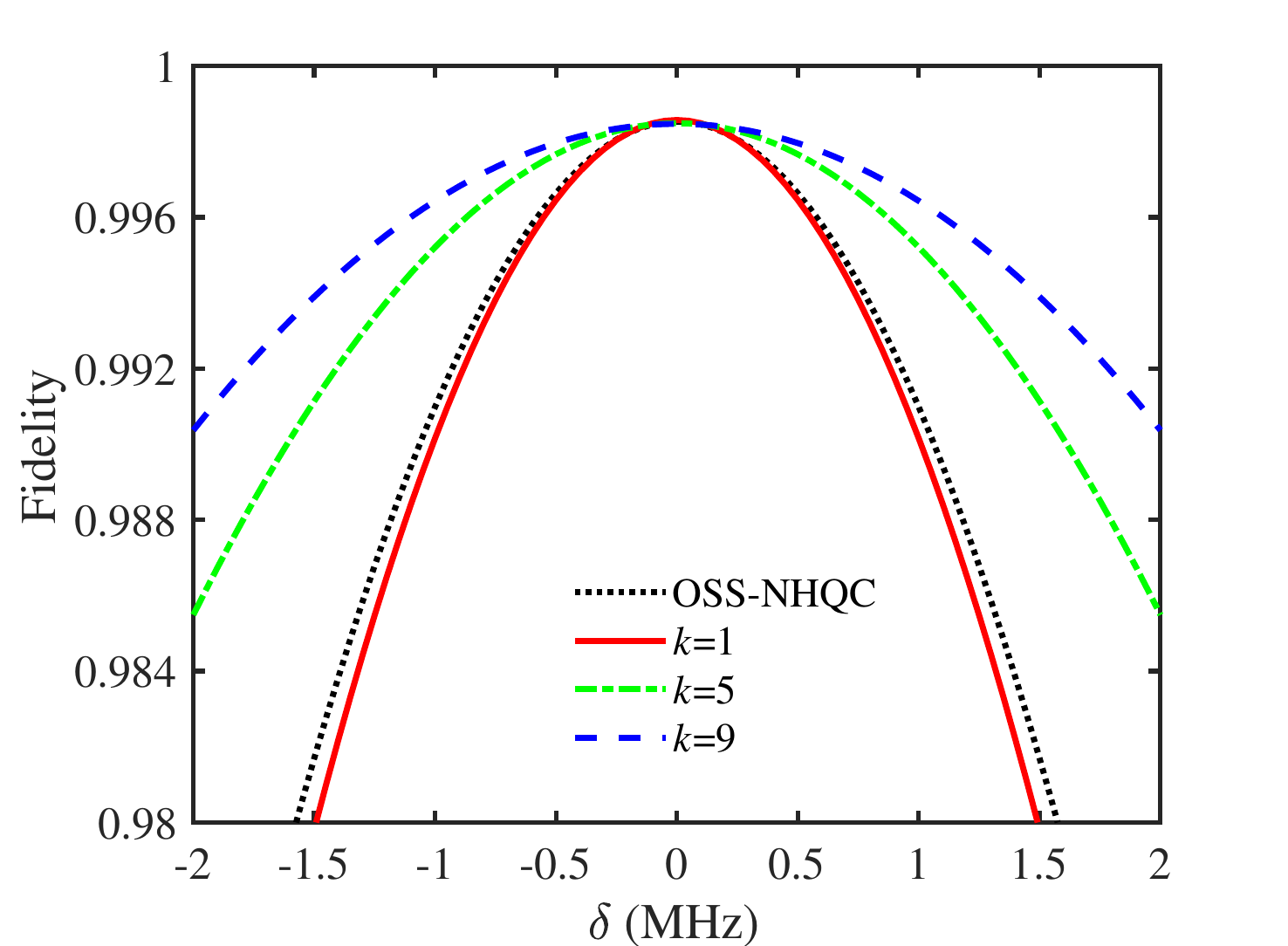}
\centerline{(d) Hadamard gate}
\end{minipage}
	\caption{(Color online)Dependence of the average fidelity on frequency detuning error for (a) T gate, (b) S gate, (c) Not gate, and (d) Hadamard gate.}
	\label{fig:Fidelity_delta}
\end{figure}

\begin{figure}[H]
\begin{minipage}{8cm}
\centering
	\includegraphics[width=7.5cm,height=5cm]{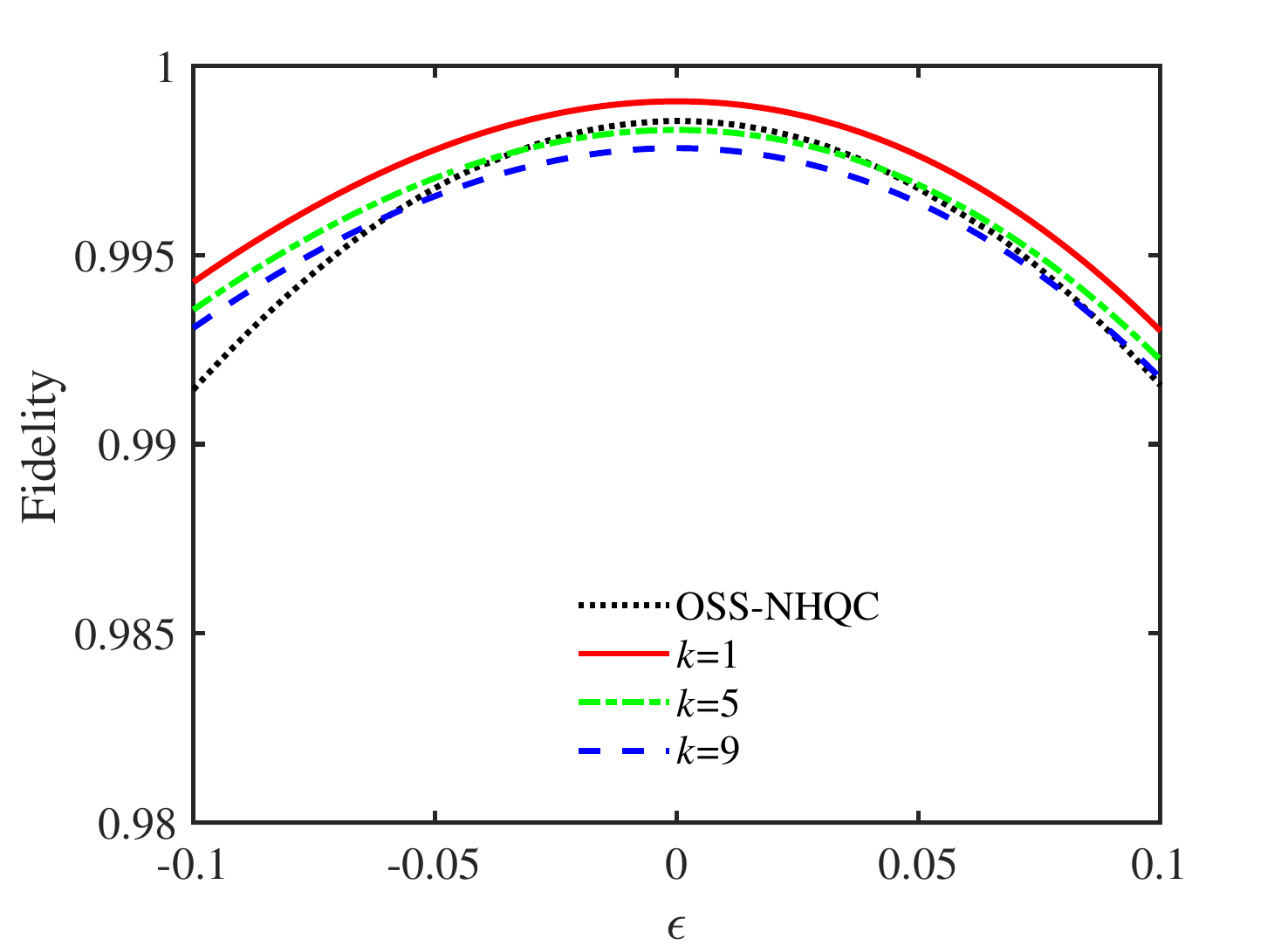} %
\centerline{(a) T gate }
\end{minipage}
\hfill  
\begin{minipage}{8cm}
\centering
	\includegraphics[width=7.5cm,height=5cm]{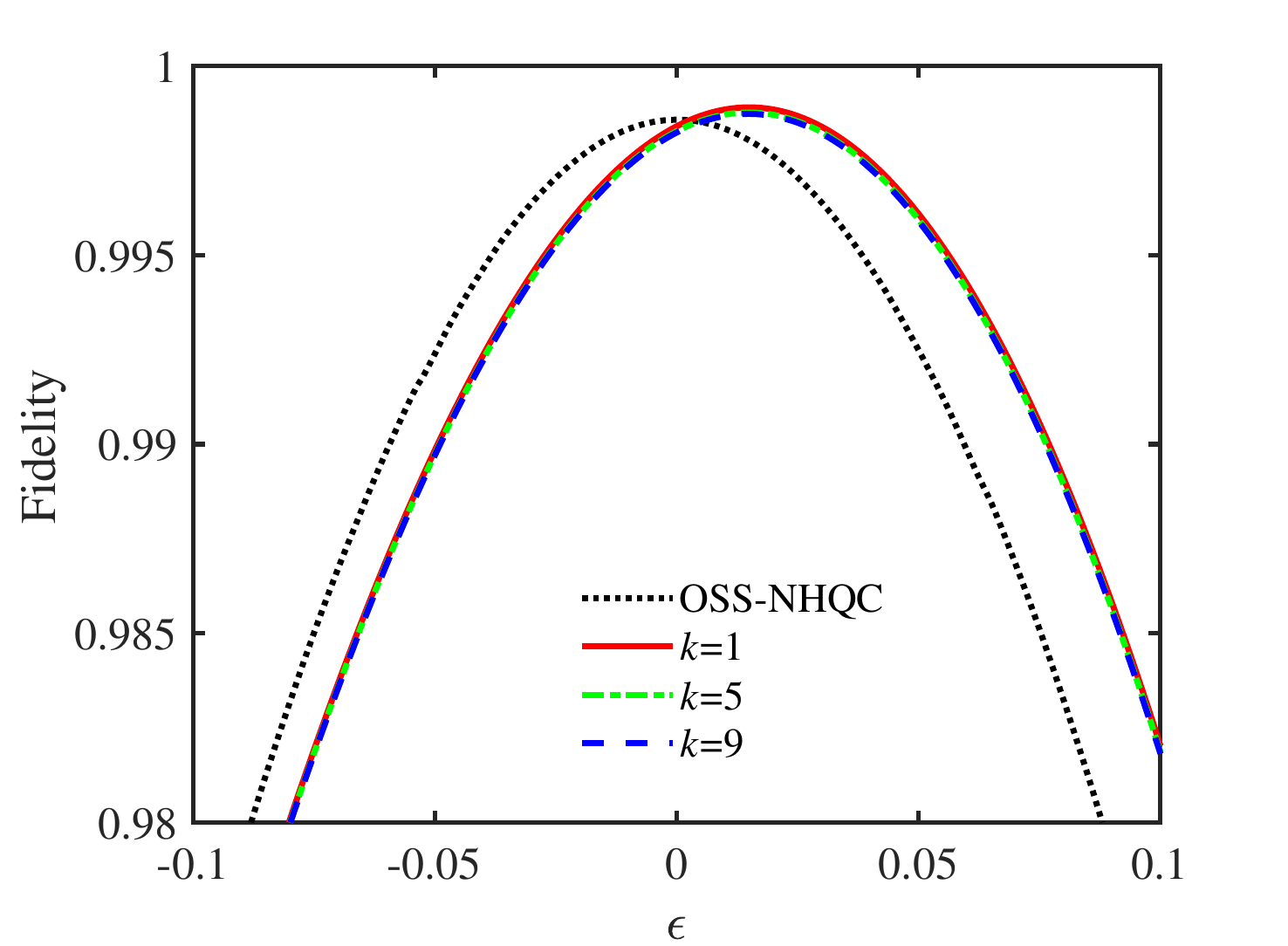}
\centerline{(b) S gate}
\end{minipage}
\vfill
\begin{minipage}{8cm}
\centering
	\includegraphics[width=7.5cm,height=5cm]{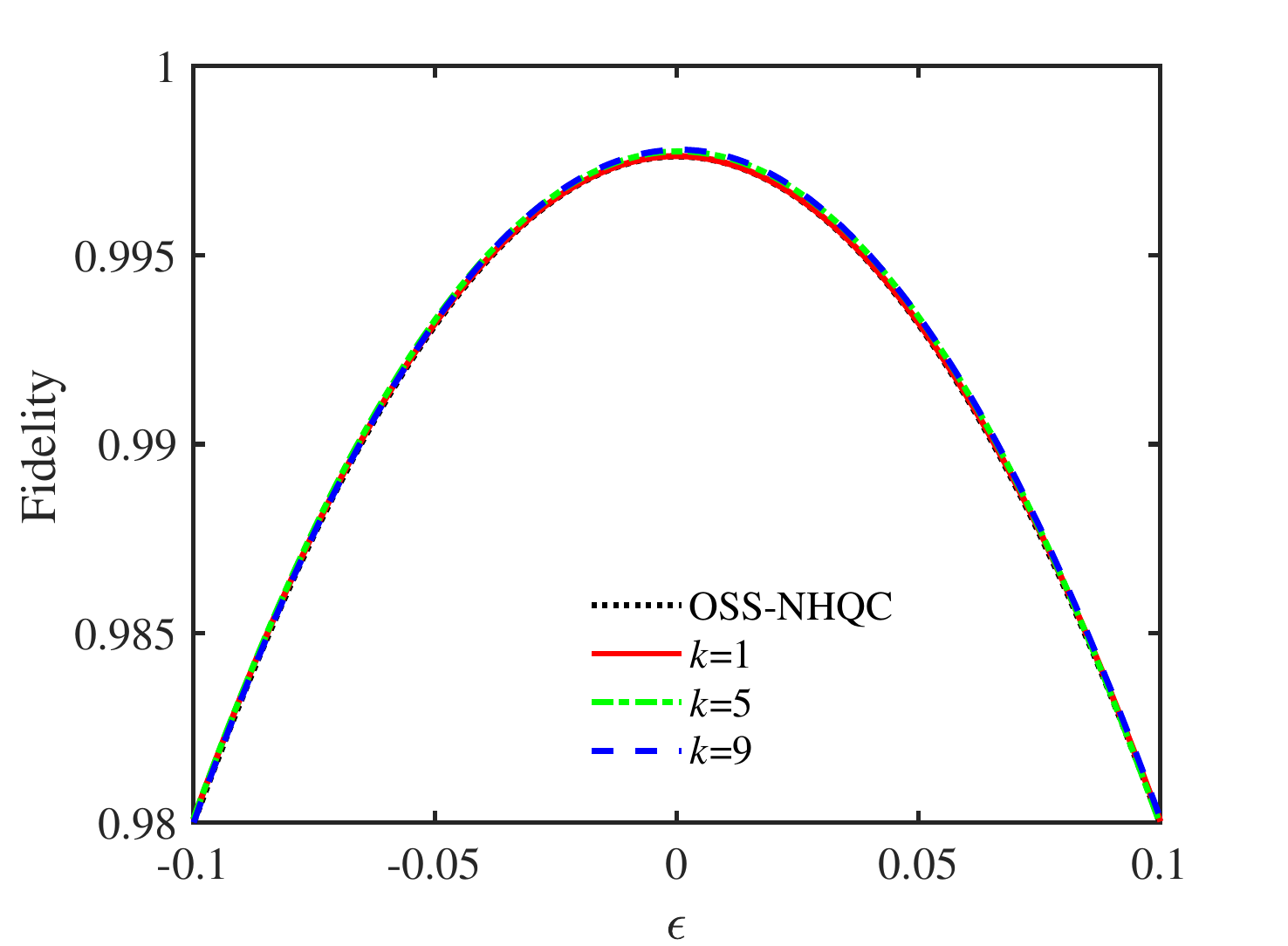}
\centerline{(c) Not gate}
\end{minipage}
\hfill
\begin{minipage}{8cm}
\centering
	\includegraphics[width=7.5cm,height=5cm]{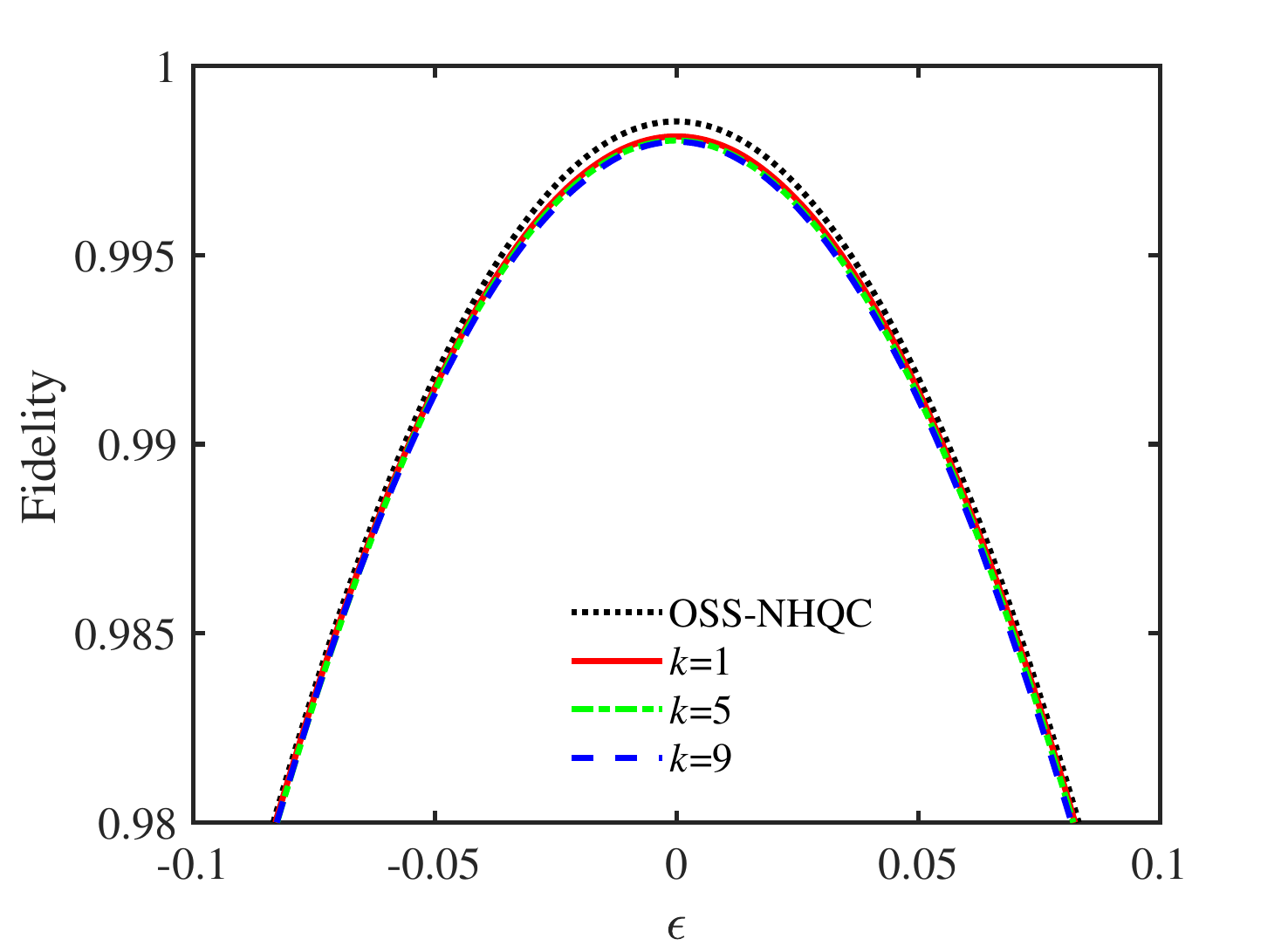}
\centerline{(d) Hadamard gate}
\end{minipage}
	\caption{(Color online)Dependence of the average fidelity on the Rabi error $\epsilon$ for (a) T gate, (b) S gate, (c) Not gate, and (d) Hadamard gate.}
	\label{fig:Fidelity_epsilon}
\end{figure}

To examine the robustness, eq.(\ref{eq:Lindblad}) is  solved numerically with the 4th-order Runge-Kutta method by varying the values of detuning error $\delta$ and  Rabi error rate $\epsilon$.
 Each average fidelity is calculated with 1001 different initial states $|\psi(0)\rangle$ which are evenly distributed with $\theta_0\in[0,\pi]$ and $\phi_0\in[0,2\pi]$. The dependence of the average fidelities on the  $\delta$ and  $\epsilon$ is shown in Figure \ref{fig:Fidelity_delta} and \ref{fig:Fidelity_epsilon} for $k=1,5,9$ in our scheme, which are plotted as solid-red, dash-dotted-green, and dashed-blue curves, respectively.  The same curve for the OSS-NHQC scheme is plotted in dotted-black. 

As we can see from Figure \ref{fig:Fidelity_delta}, the T gate and S gate in our scheme are more robust against the detuning error $\delta$ than those in the OSS-NHQC scheme. But for Not gate and Hadamard gate, the robustness curves of OSS-NHQC scheme almost coincide with the $k=1$ case, since their pulse envelops are almost the same. The advantage  in robustness for the T gate and S gate in our scheme can be explained in the framework of time-dependent perturbation theory, where the integrated perturbative effects are smaller because of the shorter evolution time. 
On the other hand, the curves with larger $k$ value in our scheme are more robust to the $\delta$.  The reason may be that the pulses with larger $k$ have higher maximum $\Omega_{\rm m}$, as we can see in Figure \ref{fig:gate_pulse}, therefore the  perturbative effect of static detuning errors $\delta$ is smaller in contrast to $\Omega_{\rm m}$. 

\begin{figure}[H]
\begin{minipage}{8cm}
\centering
	\includegraphics[width=7.5cm,height=5cm]{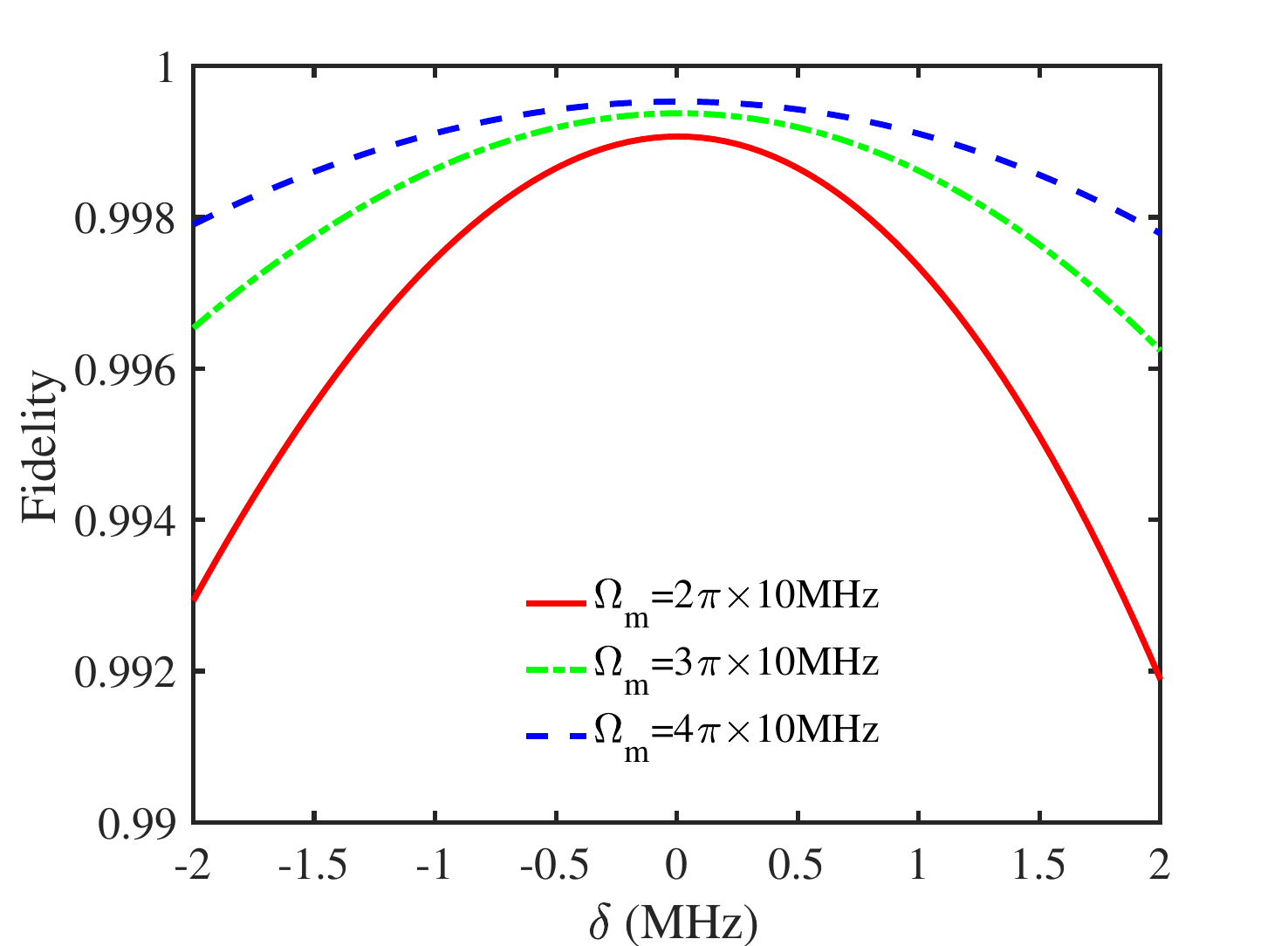} %
\centerline{(a) T gate }
\end{minipage}
\hfill  
\begin{minipage}{8cm}
\centering
	\includegraphics[width=7.5cm,height=5cm]{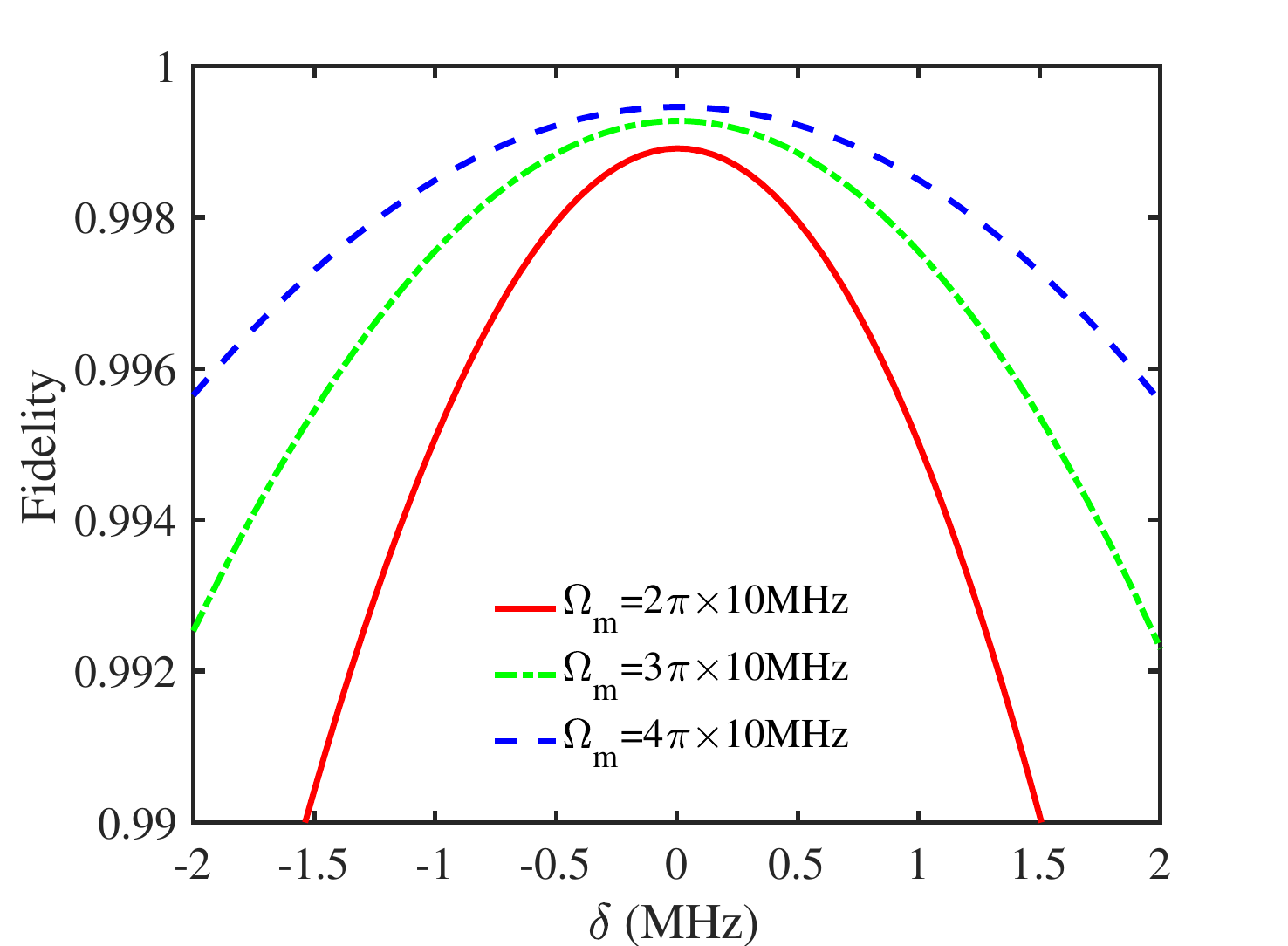}
\centerline{(b) S gate}
\end{minipage}
\vfill
\begin{minipage}{8cm}
\centering
	\includegraphics[width=7.5cm,height=5cm]{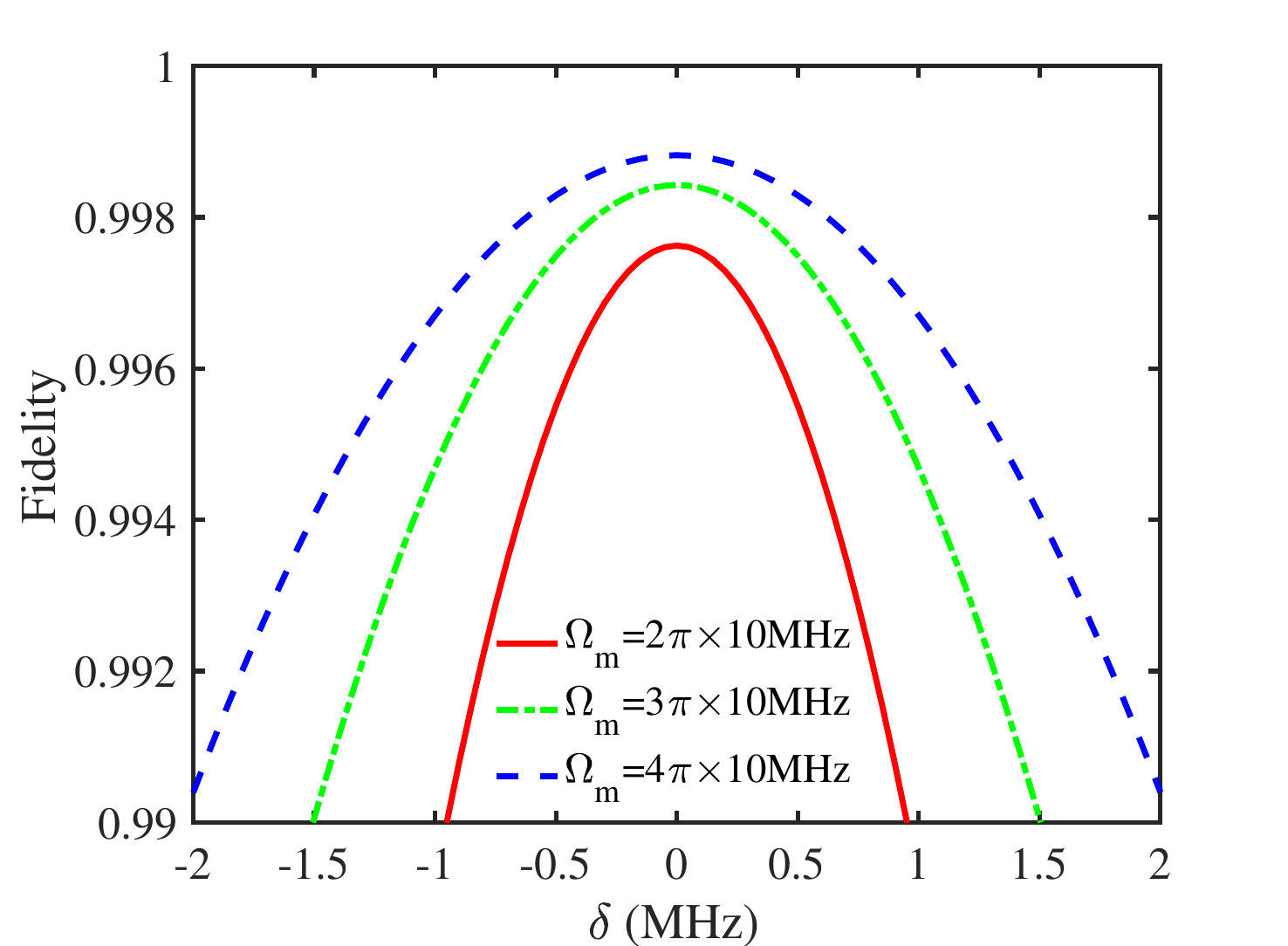}
\centerline{(c) Not gate}
\end{minipage}
\hfill
\begin{minipage}{8cm}
\centering
	\includegraphics[width=7.5cm,height=5cm]{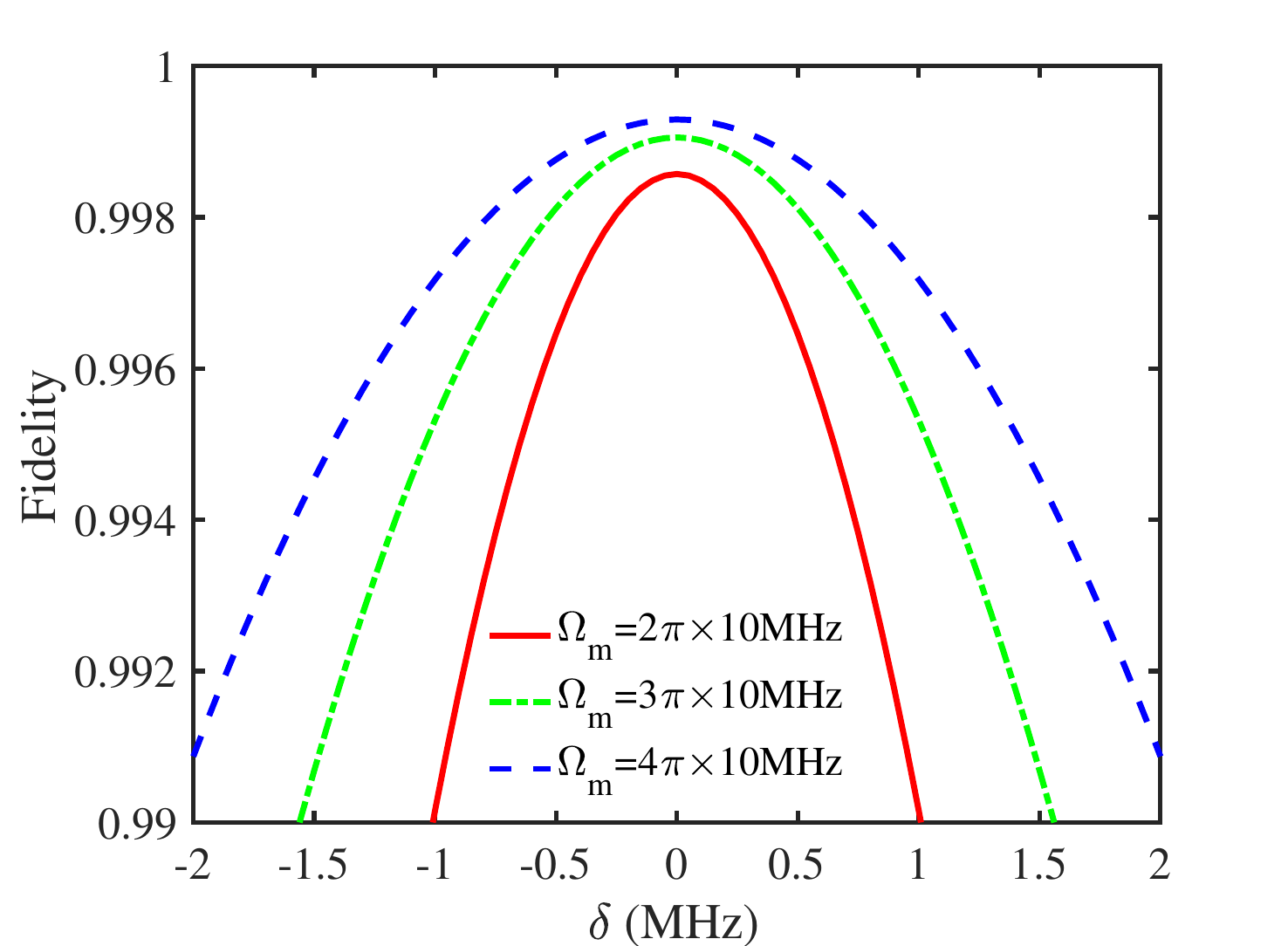}
\centerline{(d) Hadamard gate}
\end{minipage}
	\caption{(Color online)Dependence of the average fidelity on the frequency detuning error for (a) T gate, (b) S gate, (c) Not gate, and (d) Hadamard gate.}
	\label{fig:Fidelity_delta_k}
\end{figure}

 In order to further investigate the above inference on the robustness against $\delta$, we plotted the similar curves in Figure \ref{fig:Fidelity_delta_k} for different maximum  Rabi frequencies $\Omega_{\rm m}=2\pi\times10$ MHz, $3\pi\times10$ MHz, and $4\pi\times10$ MHz while keep $k=1$. These plots clearly show that the pulses with higher maximum $\Omega_{\rm m}$ are more robust against $\delta$. That is to say, the robustness of pulses against the static detuning $\delta$ is dominated by $\Omega_{\rm m}$. The higher $\Omega_{\rm m}$, the more robust against $\delta$. 

As for the robustness against the Rabi error rate $\epsilon$, there are no significant difference for different $k$ values in our scheme, as well as those in the OSS-NHQC scheme, as shown in Figure \ref{fig:Fidelity_epsilon}. This is because that the Rabi error $\epsilon\Omega(t)$ {\color{blue} defined here} is as proportional to $\Omega(t)$, so the pulse envelop is irrelevant.

\section{Discussion and Conclusion }
\label{sec:conclusion}
In this work, we proposed a method to implement an arbitrary single qubit gate of geometric quantum computing in a single-shot manner in a three-level quantum system with the $\Lambda$ configuration. Comparing with  the previous NHQC and GQC schemes whose paths are the orange-sliced shape on the Bloch sphere, our method realized a circular trajectory, which is the shortest in length. Under the condition of the parallel transport, the dynamical phase is removed, and we find that the Rabi frequency $\Omega(t)$ and the detuning $\Delta(t)$ is proportional to each other by a constant value, which is determined by the geometric phase $\gamma$. This means that the central frequency $\nu(t)$ of the laser has to be time dependent for $\gamma\neq\pi$ in general. On the other hand, this relation also leads to a square pulse $\Omega(t)=\Omega_0$ if the detuning $\Delta$ is a constant, which is the case in Ref.\cite{Erik2016}. In addition, our method reduces to the original NHQC scheme \cite{Erik2012} in the limit of $\gamma=\pi$.

We parametrized the pulses based on the polynomial of parabolic function $\bar{\alpha}(t)^{k+1}$, so that the cyclic evolution condition and the condition $\Omega(0)=\Omega(\tau)=0$ are satisfied automatically. In order to meet the condition in eq.(\ref{cyclic_omega}), smaller $k$ value implies shorter gate duration $\tau$, provided that the maximum of $\Omega(t)$ is fixed. And the lower bound of $\tau$ is determined by the square pulse in eq.(\ref{eq:tau_gamma_constant}). Generally, the gates duration with small $k$ value, e.g. $k=1$, in our scheme are shorter than those in the previous work.

We also investigated the robustness of the pulses against the static detuning error $\Delta(t)\to\Delta(t)+\delta$ and  the Rai error $\Omega(t)\to\epsilon\Omega(t)$ . Along with the OSS-NHQC scheme, three representative $k$ values ($k=1,5,9$) in our scheme are chosen to obtain the pulses for T gate, S gate, NOT gate, and Hadamard gate. The quantum evolution is numerical evaluated by the Lindblad master equation for the superconducting qubit gates. The results show that the  evolution with shorter duration is more robust against the static detuning error $\delta$, since the integrated perturbative effect is smaller. We also found that the pulses with larger $k$ values are more robust against the $\delta$. The reason behind this is that larger $k$ generates higher maximum $\Omega_{\rm m}$, so the perturbative effect of $\delta$ is smaller in contrast. On the other hand, the pulses with different $k$ values are not sensitive to the Rabi error, because it is by definition  proportional to the Rabi frequency $\Omega(t)$.

Our results clearly explain that how the pulses with fast evolution can be parametrized in the framework of non-adiabatic geometric quantum computing. This may open more possibilities to achieve more fast and robust quantum gates.

%


\section*{ACKNOWLEDGEMENTS} 
We thank Dr. Bao-Jie Liu for helpful discussion on related NHQC works.
The work was supported by National Natural Science Foundation of China (grant numbers 61505133), self-determined Research Project of the Key Lab of Advanced Optical Manufacturing Technologies of Jiangsu Province(ZZ2109).

\bibliographystyle{prsty}

\end{document}